\documentclass[a4paper,11pt]{article}
\pdfoutput=1 

\usepackage{lineno}
\usepackage{setspace}
\usepackage{multirow}

\usepackage{jinstpub} 
\title{Particle Flow with a Hybrid Segmented Crystal and Fiber Dual-Readout Calorimeter}

\author[a,1]{M.T.~Lucchini,\note{Corresponding author.}}
\author[b]{L.~Pezzotti,}
\author[c]{G.~Polesello,}
\author[d]{and C.G.~Tully}

\affiliation[a]{INFN and University of Milano-Bicocca, Milano, Italy}
\affiliation[b]{CERN, Geneva, Switzerland}
\affiliation[c]{INFN, Sezione di Pavia, Pavia, Italy}
\affiliation[d]{Princeton University, Princeton, New Jersey, USA}
\emailAdd{marco.lucchini@cern.ch}

\abstract{
In the reconstruction of physics events at future e$^+$e$^-$ colliders the calorimeter design has a crucial role in the overall detector performance. The reconstruction of events with many jets in their final state sets stringent requirements on the jet energy and angular resolutions. The energy resolution for jets with energy of about 45 GeV is required to be at the 4-5\% level to enable an efficient separation of the W and Z boson invariant masses. We demonstrate in this paper how such a performance can be achieved by exploiting a particle flow algorithm tailored for a hybrid dual-readout calorimeter made of segmented crystals and fibers. The excellent energy resolution and linearity of such calorimeter for both photons and neutral hadrons ($3\%/\sqrt{E}$ and $26\%/\sqrt{E}$, respectively), inherent to the homogeneous crystals and dual-readout technological choices, provides a powerful handle for the development of a new approach for particle identification and jet reconstruction. 
While the dual-readout particle flow algorithm (DR-PFA) presented in this paper is at its early stage of development, it already demonstrates the potential of a hybrid dual-readout calorimeter for jet reconstruction by improving the jet energy resolution with respect to a calorimeter-only reconstruction from 6.0\% to about 4.5\% for 45 GeV jets.
}


\begin{document}

\maketitle

\flushbottom
\section{Introduction}

The construction of an electron-positron (e$^+$e$^-$) Higgs factory, such as the FCC \cite{FCC_CDR}, the CEPC \cite{CEPC_CDR_Vol1}, the ILC \cite{ILC_TDR} or C$^3$ \cite{CoolCopperCollider}
, is one of the highest priorities in the mid-term future of particle accelerators with the objective of accurately measuring the properties of the Higgs boson and hunting for new physics signals. 
Identification of the key requirements for the detectors that will be recording collisions at such accelerators has started and suggests the need of high resolution for the measurement of jets to reconstruct events with Z, W or H bosons decaying in their hadronic mode \cite{azzi2021exploring,ManqiRuan_2019}. 
At the same time a detector with high energy resolution for photons enhances the potential of heavy flavor physics studies with low energy photons in their final state \cite{RoyAleksan} and improves the resolution of the $Z\rightarrow ee$ recoil mass in Higgsstrahlung events \cite{Lucchini_2020}.
In this context, the calorimeter system plays a crucial role and several technological options, capable to address these challenges, are being explored \cite{Aleksa:2021ztd}.

One of the most common approaches to achieve the required jet energy resolution of about $30\%/\sqrt{E}$, (i.e. 4-5\% for 45 GeV jets) is to exploit particle flow algorithms designed for high granularity sampling calorimeters consisting of alternating layers of tungsten, as absorber, and silicon pads as active elements.
An example of such a calorimeter which is under development is the CMS HGCAL \cite{HGCAL_TDR} which capitalizes on years of R\&D by the CALICE Collaboration \cite{Calice_collaboration_2008}. 

Such calorimeters, however, have a poor resolution for single particles, typically about $25\%/\sqrt{E}$ for EM particles and $55\%/\sqrt{E}$ for hadrons. Their performance in jet reconstruction thus strongly relies on the capability to exploit the momentum measurement of charged particles provided by a silicon tracker, leaving to the calorimeter system the measurement of photons and neutral hadrons.
Ultimately, the limitation to the jet energy resolution comes from the so-called \emph{confusion term}, i.e. the mis-association of calorimeter hits from neutral hadrons to a charged track (and vice-versa) due to the unavoidable overlap of neighboring hadron showers in the calorimeter. A highly granular adaptation of noble liquid calorimeters \cite{Aleksa:2021ztd,CERN-LHCC-96-041} for the electromagnetic compartment could also be considered to improve the energy resolution for EM particles to about $10\%/\sqrt{E}$.

A completely different approach is taken for the IDEA detector concept, consisting of a monolithic calorimeter made of scintillating and Cerenkov fibers inserted into an absorber structure which thus features a fine transverse granularity but no longitudinal segmentation \cite{Antonello2020DualreadoutCA, Antonello_2020}. By exploiting the dual-readout method \cite{RevModPhys.90.025002} such a detector is capable to achieve a resolution for single neutral hadrons of $25-30\%/\sqrt{E}$ and can reconstruct jets of 45 GeV with an energy resolution of about 6.0\% \cite{LPezzotti_thesis}.

In the following we propose an innovative approach that aims at combining the two methods above, i.e. high resolution for hadrons exploiting the dual-readout technique with a particle flow approach enabled by the addition of a segmented crystal calorimeter in front of the solenoid.
While the overall concept and potential implementation of such calorimeter was previously described in \cite{Lucchini_2020}, we focus in this paper on the development of a dual-readout particle flow algorithm (DR-PFA) and demonstrate the potential of this approach to achieve the desired jet energy resolution in the range of about 4.5-3.0\% for jet energies of 45-100 GeV.

\section{Dual-readout particle flow approach in a coarsely segmented calorimeter}\label{sec:intro}

Typical particle flow algorithms are designed to work with calorimeters that have a fine longitudinal segmentation, i.e. 20-30 layers for the ECAL and 40-50 layers for the HCAL section. The ideal absorber for such detectors is tungsten to keep the shower spatial development as compact as possible. A reasonable transverse granularity was demonstrated to correspond to about half the Moli\`ere radius in~\cite{THOMSON200925}.

Two examples of such algorithms are the \texttt{ARBOR} algorithm \cite{ruan2014arbor}, exploiting the tree-like topology of showers and currently used for CEPC simulations, and the \texttt{PandoraPFA} algorithm \cite{THOMSON200925}, which currently achieves the best performance among other similar algorithms.
Both algorithms rely on the high granularity of the calorimeter, both transversely and longitudinally, and exploit mainly (if not solely) the topological information of hits (rather than their energy) to reconstruct showers and possibly match them to a charged track.

In a calorimeter, as the one described in Section~\ref{sec:calo_description}, which features only two longitudinal layers in the electromagnetic calorimeter (ECAL) and one longitudinal layer in the hadron calorimeter (HCAL) a different approach to particle flow should be used. 

\clearpage 
\noindent
Some of the key features of the hybrid segmented crystal and fiber dual-readout calorimeter compared to typical Si-W high granularity calorimeters are:
\begin{itemize}
\item the energy resolution for photons and electrons is better by about a factor 10: \\$3\%/\sqrt{E}$ for the ECAL (instead of $25-30\%/\sqrt{E}$);
\item the energy resolution for neutral and charged hadrons is better by a factor 2-2.5: \\$25-30\%/\sqrt{E}$ for the HCAL (instead of $55-60\%/\sqrt{E}$);
\item the level of longitudinal segmentation is substantially reduced: \\3-5 total (2-4 layers in the ECAL and a single layer in the HCAL) instead of $\sim40$ ($\sim20$ layers in the ECAL and $\sim20$ in the HCAL);
\item the transverse granularity in the ECAL and HCAL is comparable to that of Si-W high granularity calorimeters.
\end{itemize}

\begin{table}[!htbp]
\centering
\caption{Comparison of calorimeter features.}
\vspace{0.2cm}
\begin{tabular}{l|c|c|c}
\hline
            	   	          &  High granularity               & Fiber-based               & Hybrid crystal      \\ 
            	   	          &  Si/W ECAL and                  & dual-readout              & and dual-readout    \\ 
                              &  scintillator based HCAL        & calorimeter               & calorimeter         \\ \hline \hline
N. of longitudinal layers     &     $> 40$                      & 1                         & 5                   \\ \hline
ECAL cell cross-section       &     25--100\,mm$^2$                      & \multirow{2}{*}{2--144\,mm$^2$}                         & 100\,mm$^2$                   \\ 
HCAL cell cross-section       &     100--900\,mm$^2$                      &                         & 400--2500\,mm$^2$                   \\ \hline
EM  energy resolution         &     $ 15-25\%/\sqrt{E}$         & $10-15\%/\sqrt{E}$        & $\approx 3\%/\sqrt{E}$                   \\ 
HAD energy resolution         &     $ 45-55\%/\sqrt{E}$         & $25-30\%/\sqrt{E}$        & $\approx 25-30\%/\sqrt{E}$                   \\ \hline
\end{tabular} 
\label{tab:calo_feat_compare}
\end{table}
\vspace{0.6cm}

What emerges clearly from Table~\ref{tab:calo_feat_compare} is that a particle flow algorithm optimized for a dual-readout calorimeter with coarse longitudinal segmentation should be designed to leverage the superior energy resolution and linearity of response to offset the effect of reduced segmentation, and thus more limited topological information on the shower.

A key asset of a dual-readout hybrid calorimeter is indeed the precise measurement of the photon and neutral hadron contributions to the jet energy which account, respectively, for about 27\% and 13\% of the total jet energy on average \cite{Lucchini_2020}.
Another advantage comes from the linearity of the calorimeter response to low energy hadrons achieved with the dual-readout correction as discussed later in Section~\ref{sec:dr_pfa}. This also leads to a Gaussian-like (tail-free) response for hadron detection which is beneficial to the final jet energy resolution (with or without a particle flow approach). Conversely, non-compensating calorimeters have, by design, an asymmetrical response to hadrons as discussed in \cite{LPezzotti_thesis}.

\newpage

\section{Description of the detector simulation}\label{sec:calo_description}

\begin{figure}[!htbp]
    \centering
    \includegraphics[width=0.99\linewidth]{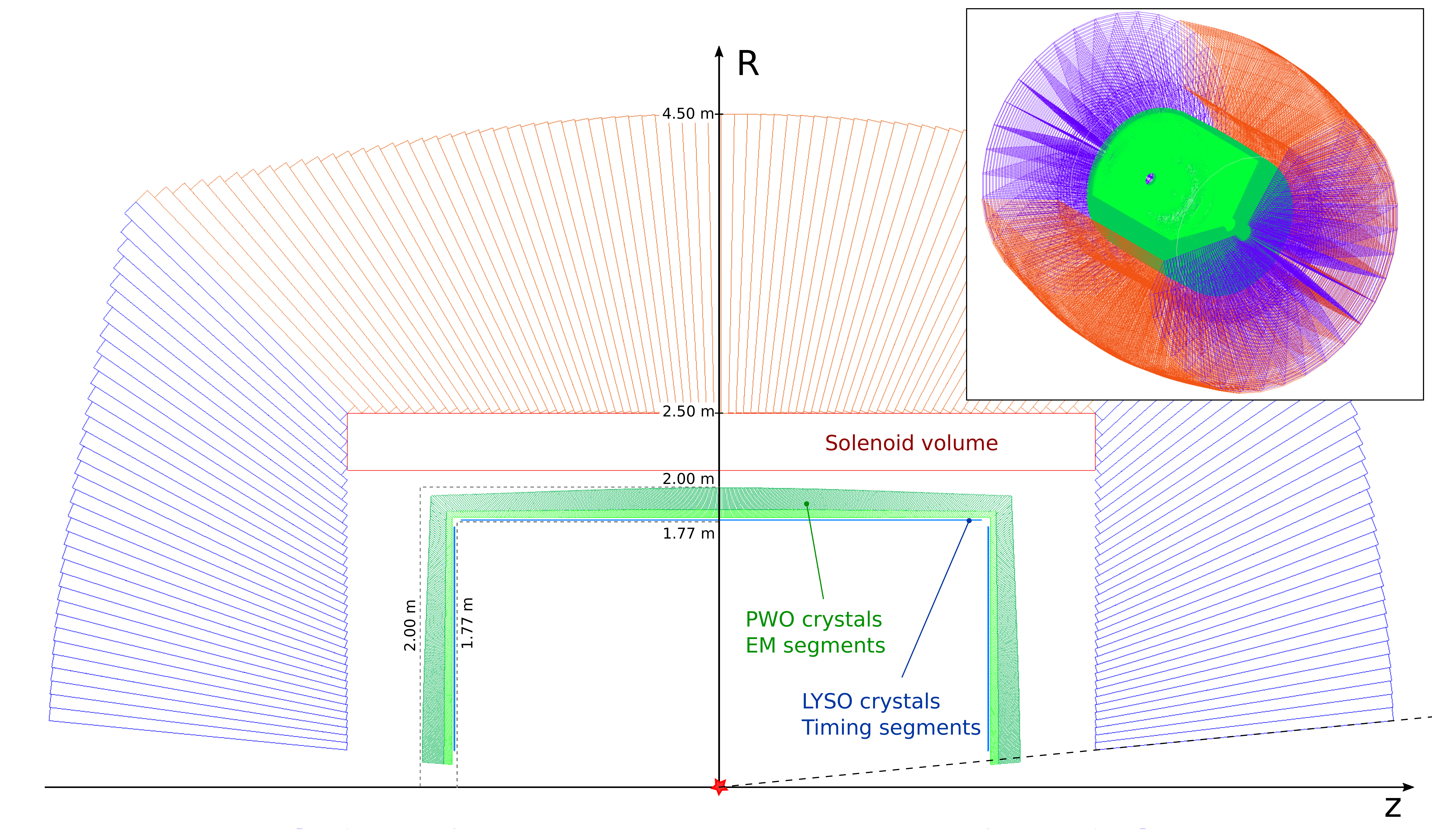}
    \caption{Side view of a single slice of the segmented dual-readout crystal calorimeter inside the IDEA solenoid (red) and fiber calorimeter (blue and orange towers).}
    \label{fig:geometry}
\end{figure}

The simulation includes a homogeneous crystal calorimeter section followed by a fiber-copper sampling calorimeter.
In the barrel region a 20 cm thick solenoid of about 0.7~$X_0$ is located between the two calorimeter segments, inspired by the ultra-thin solenoid design presented in \cite{9350162}. The radial envelopes are described in Table~\ref{tab:calo_radial_envelopes} and a picture of the simulated geometry is shown in Fig.~\ref{fig:geometry}. It should be noted that no material budget upstream the crystal calorimeter has been simulated (since its impact on the e.m. resolution was estimated in \cite{Lucchini_2020} to be negligible), however the simulation of a real tracking system is discussed as a future step in Sec.~\ref{sec:discussion}.

\begin{table}[!htbp]
\centering
\caption{Radial envelopes of the hybrid calorimeter segments.}
\vspace{0.2cm}
\begin{tabular}{l|c|c}
\hline
            	   	            &  $R_{in}$ [mm]          & $R_{ext}$ [mm]   \\ \hline \hline
Crystal timing layers           &     1775                & 1795          \\ \hline
First crystal ECAL segment     &     1800                & 1855          \\ 
Second crystal ECAL segment    &     1855                & 2000          \\ \hline
Solenoid                        &     2118                & 2500           \\ \hline
Fiber spaghetti HCAL segment    &     2500                & 4500          \\ \hline
\end{tabular} 
\label{tab:calo_radial_envelopes}
\end{table}
    
\vspace{0.4cm}

\subsection{Calorimeter geometry}
The crystal calorimeter consists of a cylindrical barrel and two endcaps, each featuring four longitudinal layers: two thin and highly segmented layers of LYSO crystals for a total of 1~$X_{0}$ dedicated to time tagging of minimum ionizing particles (and early photon conversions) with a time resolution of about 20 ps; two thicker layers of lead tungstate (PWO) crystals, a front one of about 6~$X_{0}$ and a rear one of about 16~$X_{0}$, providing a good longitudinal containment of the electromagnetic showers (total of 22~$X_{0}$).

The timing section consists of modules of thin LYSO crystal fibers with square cross section. The rear and front modules are rotated by 90 degrees with respect to each other to create a grid with granularity of $3\times3$~mm$^2$ and spatial resolution to charged particles better than a millimeter. Each crystal fiber is read out at both ends by Silicon Photomultipliers (SiPMs), in a way similar to the CMS MIP Timing Detector \cite{CMS_MTD_TDR}.

The electromagnetic calorimeter layer is made of two lead tungstate (PWO) crystals with roughly $1\times1$~cm$^2$ cross section. One SiPM is used to read out the signal from the front side of the front crystal and two SiPMs on the rear side of the rear crystal. The two SiPMs on the rear crystal are used to provide dual-readout information for hadrons that start showering in the crystals. 
The barrel ECAL is located inside the solenoid of the IDEA detector and contained in a radial envelope of $1800<R<2000$~mm. Its transverse segmentation is defined to provide a cross section of each crystal of about 1~cm$^2$, thus featuring 1130 rotations around the beam axis and 360 ($2\times180$) crystals along the angle $\theta$. Both layers of crystals have a fixed total length of $22~X_{0}$ (about 0.97$\lambda_I$) and are arranged in a projective geometry.
Each endcap consists of two truncated cones pointing to the interaction point and divided into 179 concentric rings along $\theta$. Each ring is segmented along $\phi$ to provide an approximate crystal cross section of $1\times1$~cm$^2$ as for the barrel.

The sampling calorimeter layer, with inner radius of 2.5~m and a depth of 2.0~m, is designed for the precise measurement of hadron showers and consists of a dual-readout, longitudinally unsegmented and fully projective, fiber calorimeter.
The active elements are scintillating (polystyrene) and clear-plastic fibers (PolyMethyl MethAcrylate (PMMA)) embedded in copper. The scintillation (S) and Cerenkov (C) fibers uniformly instrument the calorimeter volume in a chess-board like geometry with a $1.5$ mm pitch. The fiber diameter is $1$ mm thick (core + cladding) so that each fiber is separated from the closest ones by $0.5$ mm of absorber material. More details are given in \cite{LPezzotti_thesis}.

\subsection{Calorimeter signals and response}

The detector response includes a number of instrumental effects such as photo-statistics fluctuations, simulated by converting the deposited energy signal into the expected number of photo-electrons (phe) and by applying the corresponding Poissonian smearing. This is performed for the signals from both crystals and fibers. Preliminary ray-tracing simulations indicate that a yield of 2000 phe/GeV and 160 phe/GeV can be achieved for the S and C signals from the crystal segment, respectively, by attaching a pair of Silicon Photomultipliers and optical filters on each crystal to simultaneously detect and separate the scintillation and Cerenkov components \cite{Lucchini_2020}. Such values are used in the following to calculate the average number of phe on which the Poissonian smearing is applied. 
The S and C signals reconstructed from the fiber calorimeter instead assume light yields of 400 and 100 phe/GeV respectively, and take into account Birks' saturation effects. Similar light yields were experimentally measured in recent beam tests with electromagnetic showers \cite{ANTONELLO201852}.

The noise contribution to the detected signal (from electronics and SiPM dark count rate) has not been simulated as its impact on the measured energy is estimated to be negligible compared to photo-statistics fluctuations. Assuming a SiPM-based readout the noise contribution to the energy resolution is about $0.2\%/E$ and $2.5\%/E$ (with E in GeV), respectively for the S and C signals from the PWO crystals \cite{Lucchini_2020} and similarly negligible for the S and C fibers where smaller signals but also smaller area SiPMs are involved.

In the following study the information from the timing segments is not used.
The granularity of the calorimeter cells in the crystal section corresponds to about 0.4 mrad in both $\phi$ and $\theta$ directions (i.e. $1\times1$~cm$^2$ cross section at the front face) while the fibers in the sampling hadron calorimeter section are grouped into $5\times5$~cm$^2$ readout towers. It should be noted that in the baseline design of the IDEA fiber calorimeter without a crystal section, the readout granularity is much finer, either at the single fiber level or by grouping fibers by arrays of $8\times8$, thus yielding less than $2\times2$~cm$^2$ wide towers. With the presence of a crystal section that measures accurately the electromagnetic particles (photons and electrons) we considered that a granularity in the HCAL section of about $5\times5$~cm$^2$ is sufficient as hadron showers have an intrinsic spread over a wider transverse region. Such level of granularity in the ECAL and HCAL sections is considered to be a reasonable compromise between particle flow performance and channel count, based on the study performed in \cite{THOMSON200925} using the PandoraPFA algorithm and a Si-W calorimeter.

\begin{figure}[!tbp]
    \centering
    \includegraphics[width=0.99\linewidth]{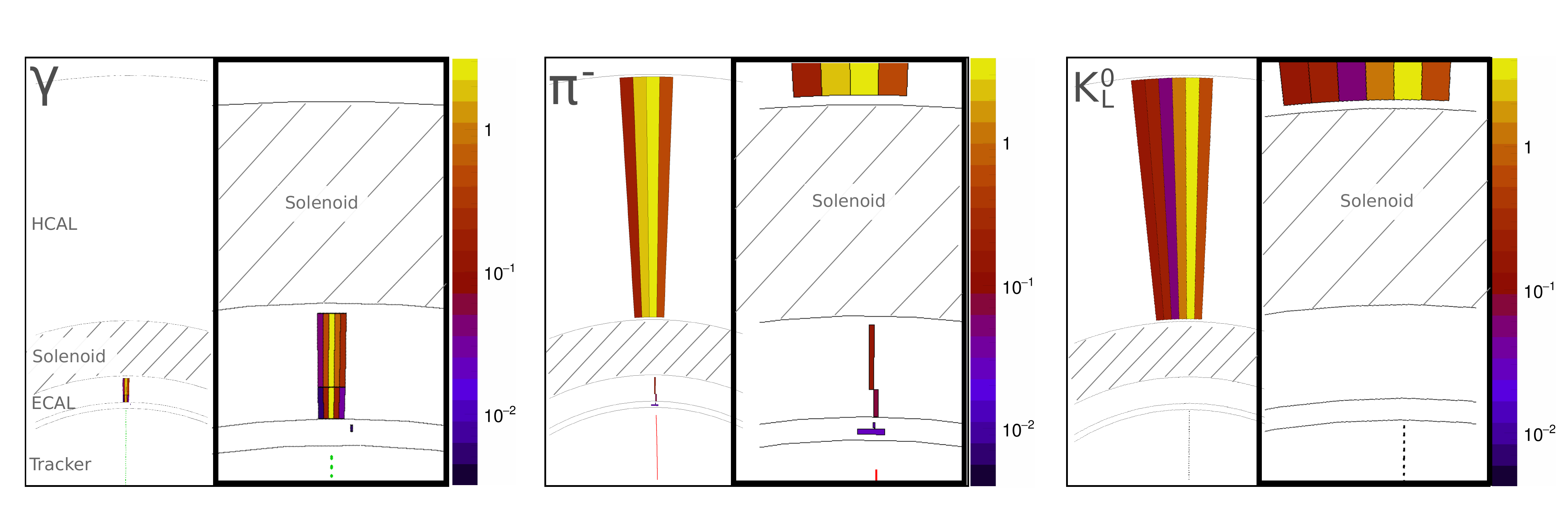}
    \caption{From left to right: single event display of a photon, charged hadron ($\pi^-$) and neutral hadron ($K^{0}_{L}$) interacting in the hybrid dual-readout calorimeter. All hits with energy above 2 MeV are displayed and the color scale is proportional to the energy of the hit.}
    \label{fig:single_particle_pfa}
\end{figure}

The detector response is stored in collections of hits, each hit containing the calibrated energy response extracted from both the scintillation and Cerenkov signal and the information on the position of the hit inside the calorimeter (x,y,z).
As an example of the information used by the DR-PFA algorithm described in the next Section~\ref{sec:dr_pfa}, event displays for a single photon, a charged hadron ($\pi^-$) and a neutral hadron ($K^{0}_{L}$) of 10 GeV are shown in Fig.~\ref{fig:single_particle_pfa}.

The response and calibration of the hadronic section of the calorimeter to single particles, both electromagnetic and hadronic ones, was described elsewhere \cite{LPezzotti_thesis}. A similar procedure has been used for the calibration of the crystal calorimeter segment. For the latter, an energy resolution of about $\sigma_{E}/E=2.5\%/\sqrt{E}\oplus 0.6\%$ and an angular resolution in mrad of $\sigma_{\theta}=1.56/\sqrt{E}\oplus 0.33$ for electromagnetic showers was obtained using the same geometry described in Section~\ref{sec:calo_description}. Response to single hadrons is reported in Section~\ref{sec:dr_pfa} as instrumental to the particle flow algorithm.  

\subsection{Simulation of tracks}
The IDEA detector concept features a tracking system consisting of a central tracking device based on ultra light drift chambers, a silicon pixel detector for the vertex region and a silicon wrapper located just in front of the calorimeter \cite{Tracker_forIDEA}. Such a detector represents a total material budget smaller than 0.02$X_0$ in the barrel region.
For the scope of this paper the tracking system was not simulated and the trajectories of particles are thus estimated based on the Monte Carlo (MC) truth information.

\begin{figure}[!tbp]
    \centering
    \includegraphics[width=0.495\linewidth]{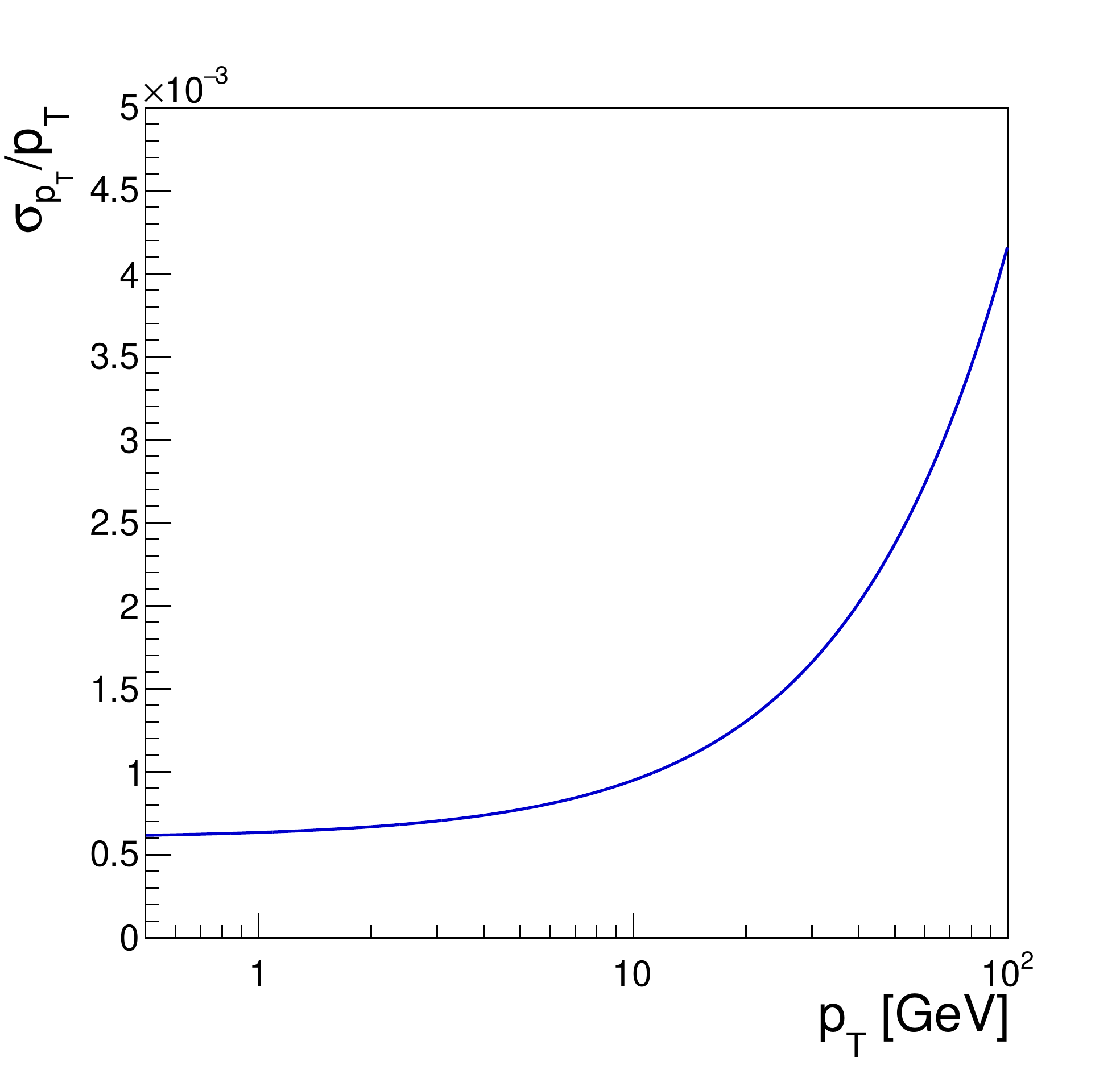}
    \includegraphics[width=0.495\linewidth]{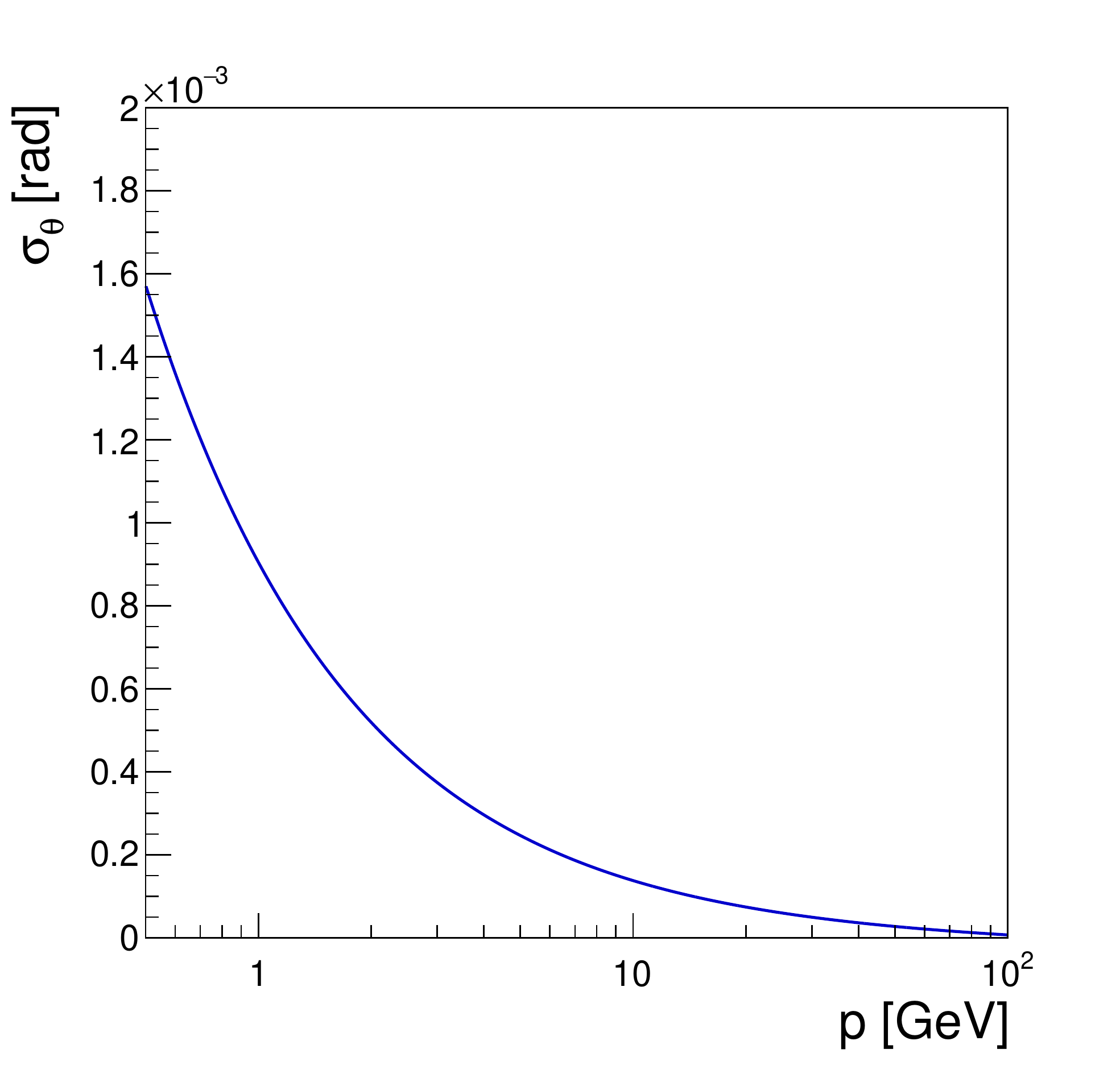}
    \caption{Parameterization of the tracker transverse momentum resolution (left) and of the $\theta$ angular resolution (right) used for smearing of Monte Carlo truth tracks according to the expected performance of the IDEA tracking system \cite{Tracker_forIDEA}.}
    \label{fig:tracker_resolution}
\end{figure}

Tracks are simulated as helical trajectories inside a uniform magnetic field of 2~T based on their momentum at the interaction vertex provided by the simulation. A Gaussian smearing is applied to the transverse momentum and direction ($\theta$) of each particle according to the expected performance of the IDEA tracker \cite{Tracker_forIDEA} described by the following parameterizations, reported for reference in Fig.~\ref{fig:tracker_resolution}, of the transverse momentum resolution (with $p_T$ in GeV/c):

\begin{equation}
\frac{\sigma_{p_{T}}}{p_{T}} = A \oplus B \cdot {p_{T}} \oplus C\cdot{p_{T}^{2}}
\end{equation}
with  $A=6\times 10^{-4}$, $B=2.024\times 10^{-4}$, $C=3.593\times 10^{-5}$ and of $\theta$ angular resolution given in radians by the formula: 
\begin{equation}
\sigma_{theta} = A\cdot p_{T}^{\alpha} \oplus B\cdot p_{T}^{\beta} \oplus C\cdot p_{T}^{\gamma}
\end{equation}
\noindent
with $A=2.75\times 10^{-6}, \alpha =0.224, B= 9.13\times 10^{-4}, \beta=-0.794, C = 1.52\times 10^{-6}, \gamma=0.384$.

\section{The Dual-Readout Particle Flow Algorithm (DR-PFA)}\label{sec:dr_pfa}
\begin{figure}[!tbp]
    \centering
    \includegraphics[width=0.99\linewidth]{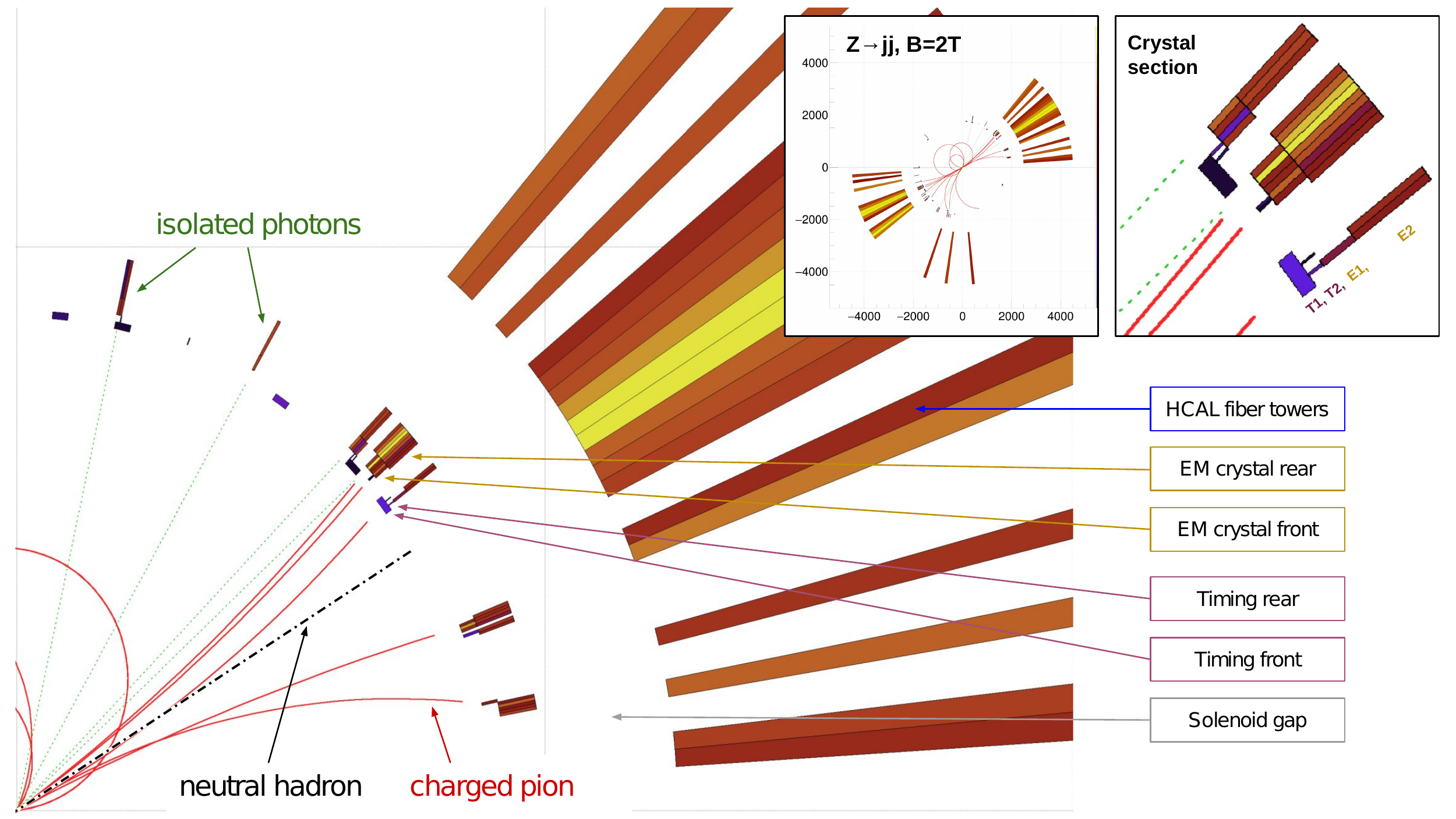}
    \caption{Event display of a jet from $Z\rightarrow jj$ event at center-of-mass energy of 90 GeV, with a 2~T magnetic field. The Monte Carlo tracks of charged pions are shown as red lines, neutral hadrons are represented as dashed-dotted black lines while photons (mainly from $\pi^0$ decays) are shown as green dashed lines. The 2 layers of the LYSO crystal timing section, the 2 layers of the PWO crystal ECAL section, the solenoid gap and the readout towers of the fiber sampling hadron calorimeter are shown from inner to outer radius.}
    \label{fig:jet_event_display}
\end{figure}

The Dual-Readout Particle Flow Algorithm (DR-PFA) has been tested and tuned using a sample of $e^+e^- \rightarrow Z/\gamma^* \rightarrow jj$ events with center-of-mass energy of 90 GeV. The two jets in the event are produced back-to-back and equally share the energy of the event so that each jet has an energy of about 45 GeV.
In this section we illustrate the outputs of the algorithm at each step and each step is described in detail.
%
\noindent
The algorithm developed consists of the following steps:
\begin{itemize}
    \item Identification of calorimeter hits belonging to photons and removal of such hits from the hit collection.
    \item Association of the remaining calorimeter hits to charged tracks by exploiting the dual-readout corrected response of the hybrid calorimeter.
    \item Clustering of successfully matched charged tracks, tracks that do not reach the ECAL surface, and unmatched calorimeter hits (ideally from photons and neutral hadrons) into jets.
    \item Application of the dual-readout correction on the fraction of jet energy from unmatched hits.
\end{itemize}
For illustrative purposes, the event display of a jet from a Z boson decay is shown in Fig.~\ref{fig:jet_event_display}.

\subsection{Photon identification}\label{sec:photon_hits}
Photons within jets are mainly the product of $\pi^0$ decays and have an energy distribution with most probable value around 1 GeV and thus can be considered entirely contained in the crystal section of the calorimeter.

A first algorithm runs over all ECAL hits and identifies \emph{neutral seeds}. A neutral seed is defined as an ECAL hit with energy larger than 100 MeV and no track impacting on the calorimeter within a radius on the $\theta-\phi$ plane, $\Delta R = \sqrt{\Delta \theta^2 + \Delta \phi^2}$, of 0.013 (corresponding to about a matrix of $4\times 4$ crystals).
Among several neighboring hits above the energy threshold, the seed is defined as the one with the highest energy in a radius of $\Delta R =0.013$.
Once a set of neutral seeds is identified in the event, all crystal calorimeter hits within a radius $\Delta R =0.013$ are clustered around each seed.

\noindent
A neutral seed cleanup step is then performed based on a simple estimator of the transverse shower shape of each seed and the corresponding hits defined according to Eq.~\ref{eq:seed_Rt}. 
\begin{equation}\label{eq:seed_Rt}
R_{transverse} = \frac{E_{seed}}{\sum_i E_{hit,i} (\Delta R_i<0.013) }
\end{equation}

\noindent
In particular, all seeds for which the ratio of the energy of the seed over the total energy of the clustered hits, $R_{transverse}$, is larger than 0.9 are excluded as they most likely belong to minimum ionizing tracks (or possibly to neutral hadrons which initiate the showering) rather than to electromagnetic showers which have typically a $R_{transverse} = 0.3-0.8$. 

\begin{figure}[!tbp]
    \centering
    \includegraphics[width=0.495\linewidth]{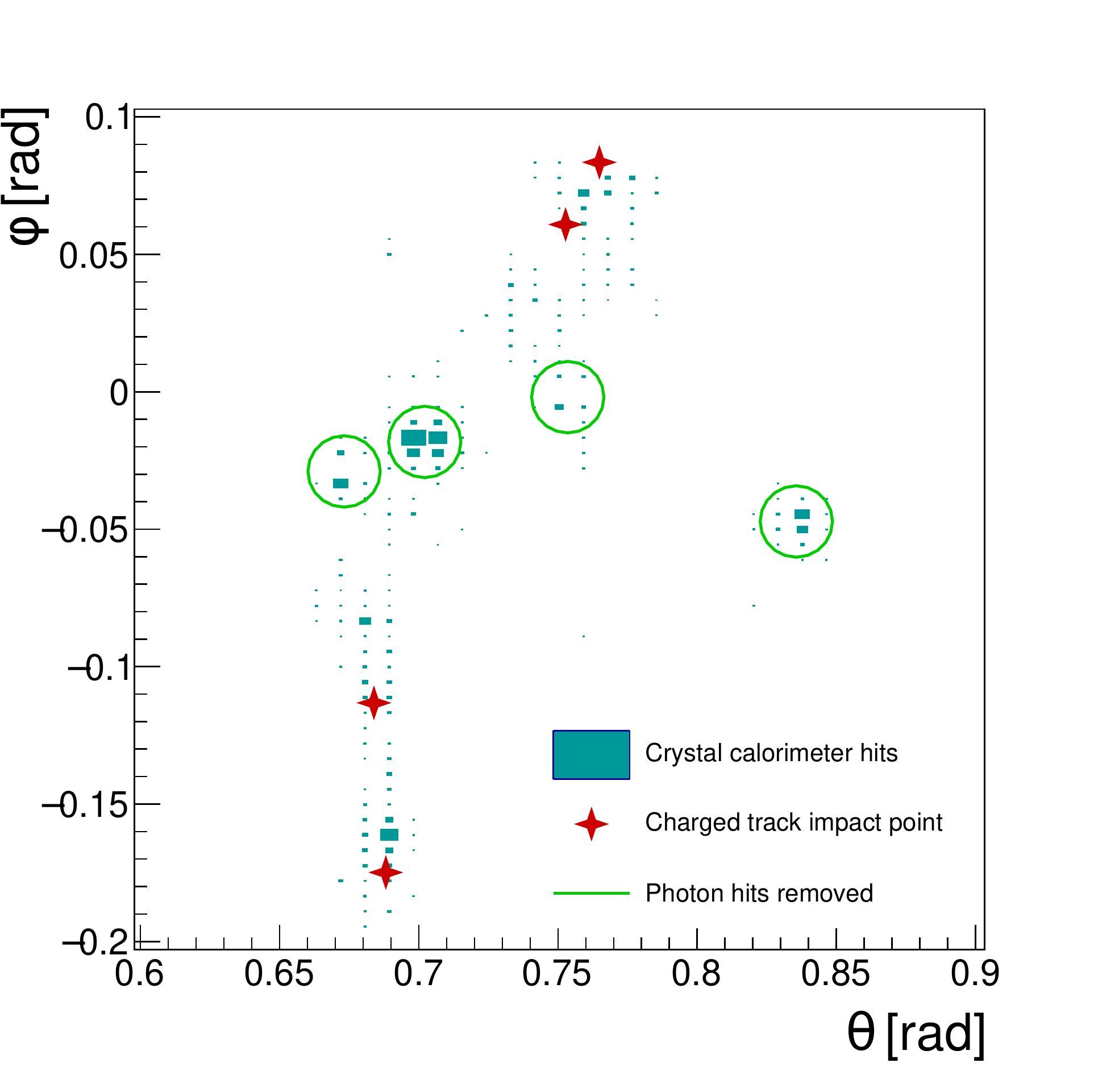}
    \includegraphics[width=0.495\linewidth]{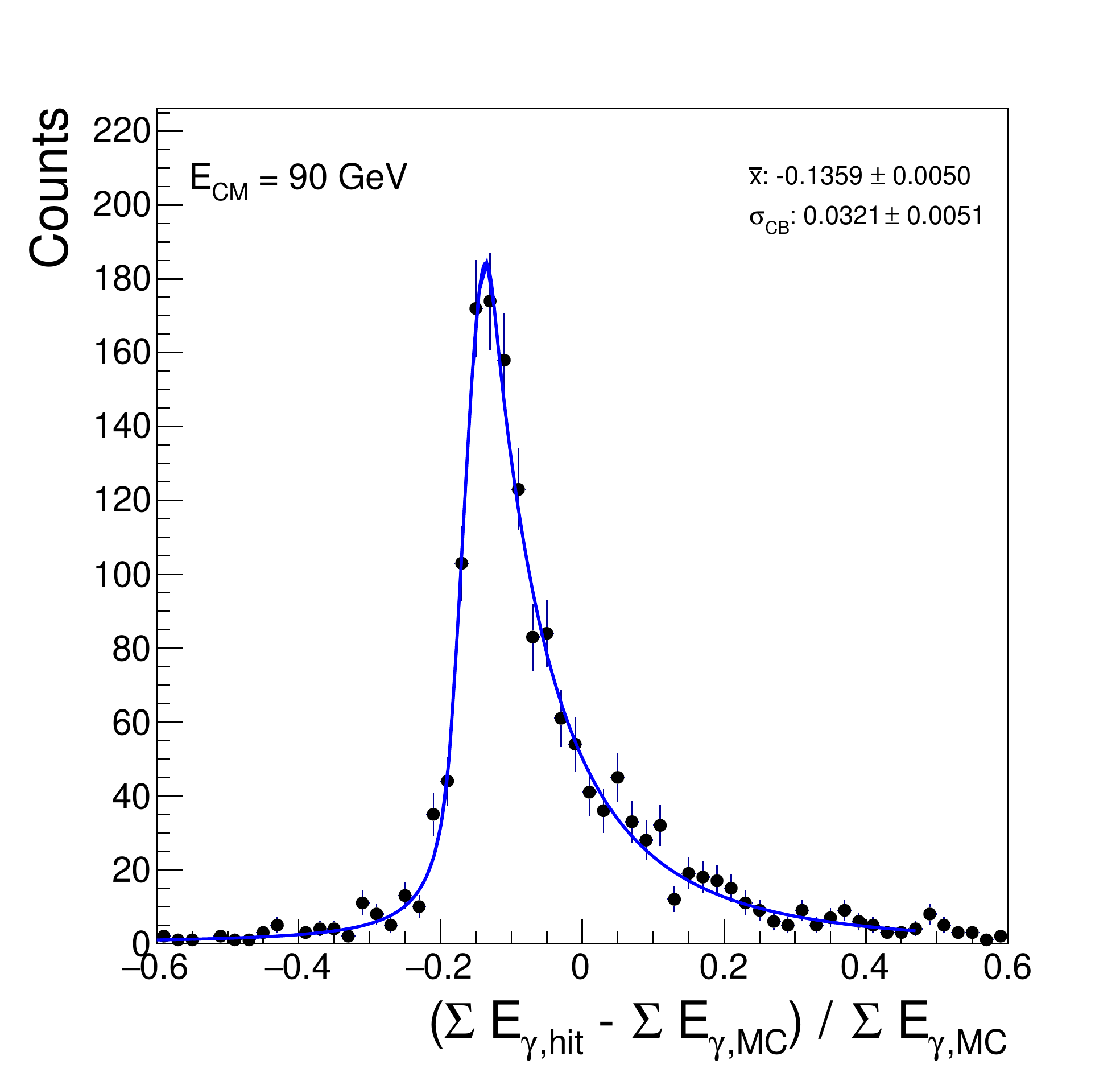}
    \caption{Left: event display of the calorimeter hits (projective sum of the two PWO segments) for a jet. Two pairs of photons are present according to MC information and the calorimeter hits within the green circle (a distance $\Delta R < 0.013$ are removed from the hit collection). The red stars show the estimated impact point of charged tracks (belonging in this case to charged pions) on the crystal calorimeter ($R=1900$~mm). Right: residual distribution of the total energy of the calorimeter hits removed at this step minus the expected energy from photons within the event.}
    \label{fig:ppfa_step1}
\end{figure}

All seeds and calorimeter hits which have been identified as belonging to neutral particles (mainly photons) by this algorithm are temporarily removed from the collection of calorimeter hits that will be fed to the next step of the algorithm and will not take part to the track-hit matching procedure.
A visual representation of this step is given in Fig.~\ref{fig:ppfa_step1}. The relative difference between the total energy of hits identified as photons and the total energy of photons in the event is shown and fitted with a Crystal Ball function. The distribution peaks at about $-0.14$ since the radius of 0.013 used for clustering hits only allows to contain about $85-90\%$ of an electromagnetic shower. The narrow peak demonstrates that the procedure is effective in removing calorimeter hits belonging to photons although the tail on the positive side shows that some overlap between charged and neutral hadron hits and photon hits can occur.

\subsection{Matching of charged tracks to calorimeter hits}
For the association of calorimeter hits to a certain track the algorithm requires the impact position of each track (on the $\theta-\phi$ plane) extrapolated at a certain depth of the calorimeter and the total energy that a certain track is expected to deposit inside the calorimeter.
The impact point of charged tracks in the middle of the crystal calorimeter ($R=1900$~mm, corresponding approximately to the shower maximum) is thus calculated and each track is assumed to have the same mass of a charged pion for calculation of the track energy (\emph{pion-mass hypothesis}). This latter assumption is a reasonable approximation since about 85\% of the tracks are charged pions, about 10\% are charged kaons and only 5\% are protons or other charged particles. By exploiting the particle identification capabilities of the detector (dE/dX measurement) \cite{Tracker_forIDEA} one could further improve the precision on the expected track energy with respect to the pion-mass approximation.

\begin{figure}[!tbp]
    \centering
    \includegraphics[width=0.492\linewidth]{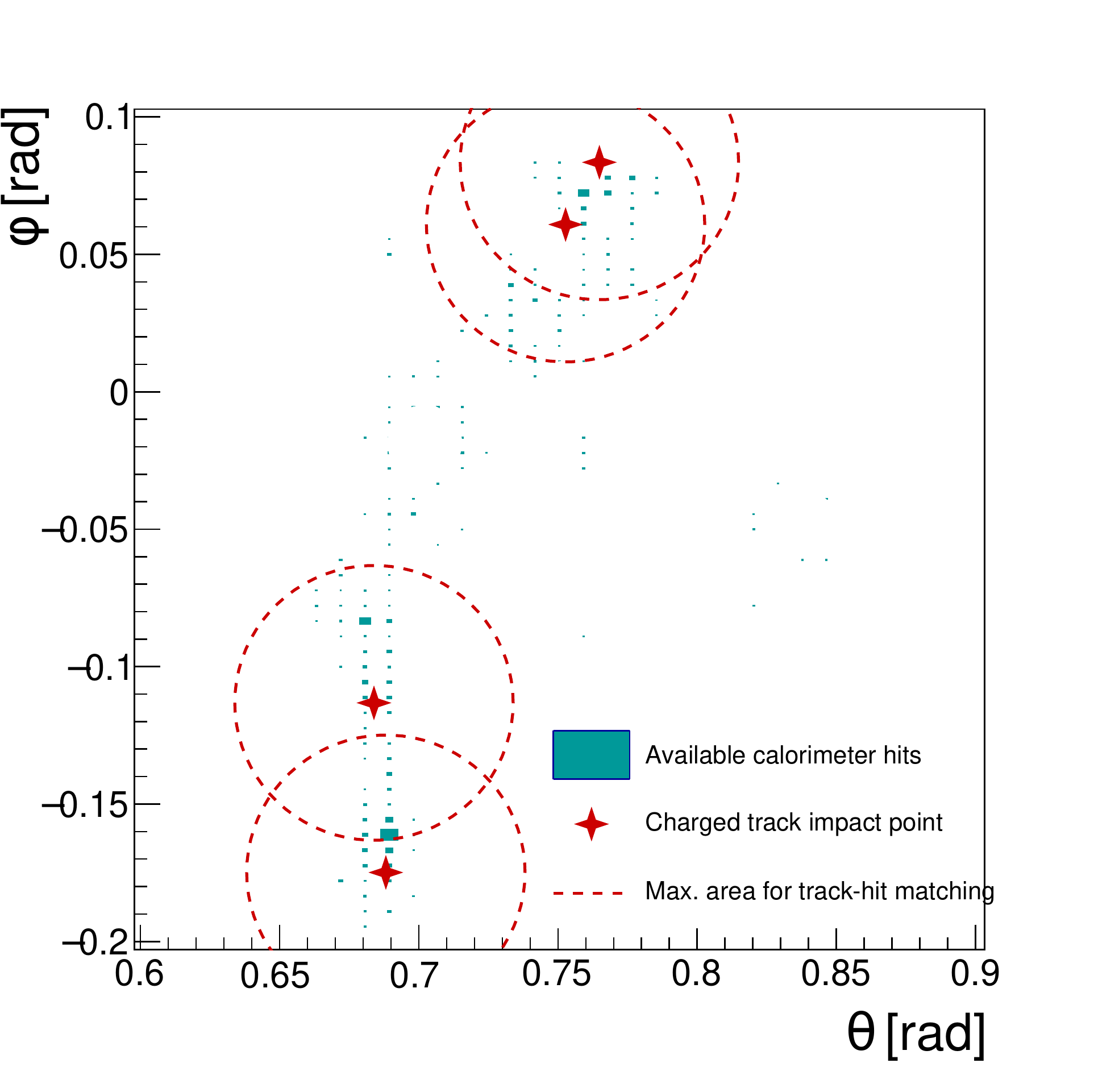}
    \includegraphics[width=0.502\linewidth]{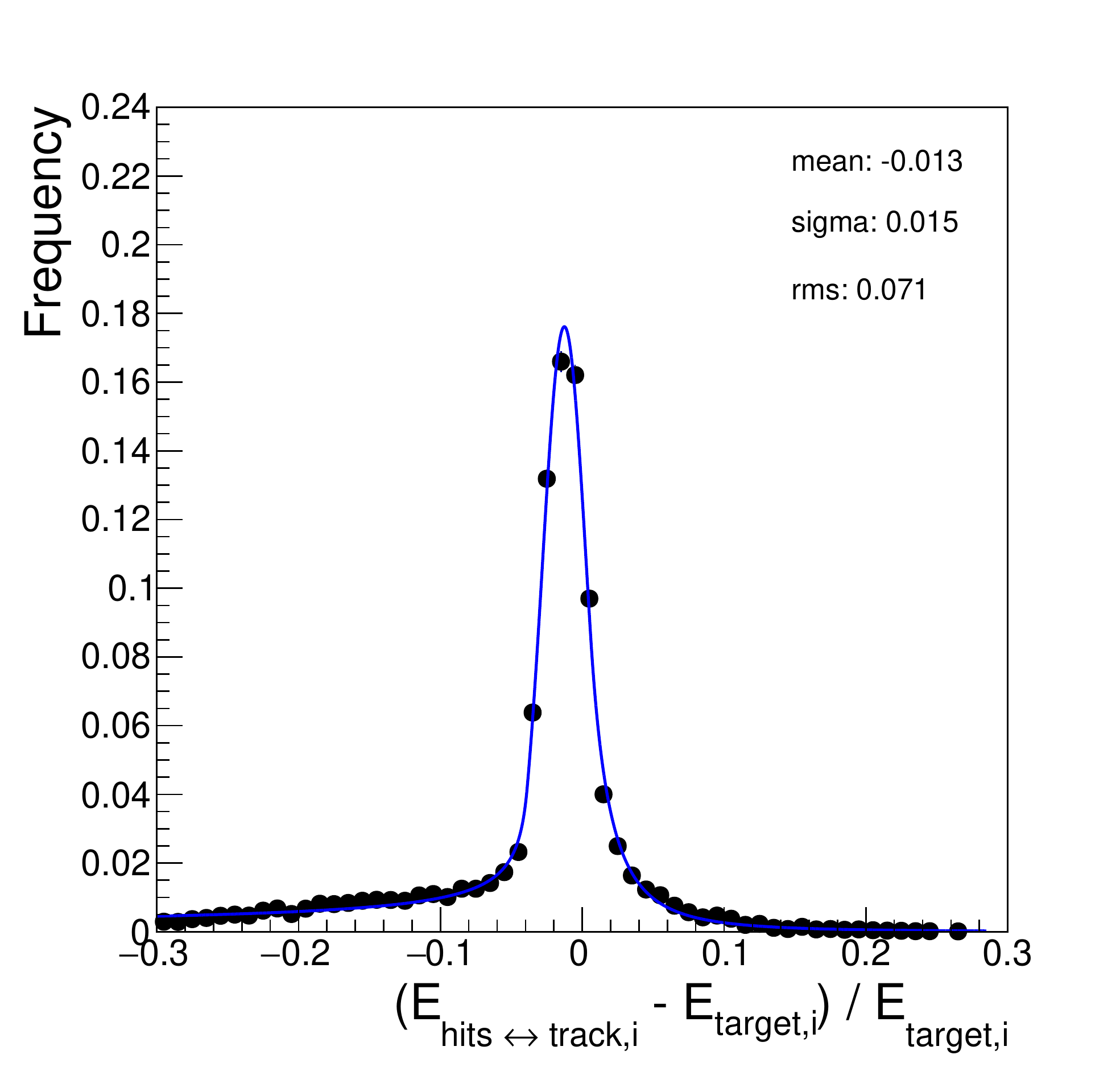}
    \caption{Left: projective sum of the crystal calorimeter hits (blue squares) on the $\theta-\phi$ plane remained after the removal of \emph{photon hits}. Projected impact points of charged tracks on the calorimeter (red stars) and circles of maximum distance $\Delta R_{max}^{ECAL} = 0.050$ (red dashed circles) which define the range for the track-hit association process. Right: relative residual between the sum of the calorimeter hits associated to a certain track and the MC smeared energy of that track.}
    \label{fig:ppfa_step2}
\end{figure}

As a first step, all charged tracks that do not reach the calorimeter are directly added to the \emph{PFA collection}: a collection of four-momentum vectors that will be used later as input to the jet clustering algorithm.
A loop is performed on the remaining charged tracks reaching the calorimeter. For each track, the collection of calorimeter hits (of both the ECAL and HCAL layers) is sorted by the distance $\Delta R$ from the track, i.e. calorimeter hits are ordered from the closest to the furthest with respect to a certain track.
The ordered collection of calorimeter hits is then parsed and hits (within a maximum cut-off distance $\Delta R_{max}$) are associated to the corresponding track as long as the addition of a new hit brings the total (dual-readout corrected) clustered energy closer to the expected calorimeter response of a certain track, $E_{target}$.
The left panel of Fig.~\ref{fig:ppfa_step2} shows the collection of calorimeter hits in the crystal segment, after removal of the photon contributions, on which the track-hit matching algorithm is run. The red dashed circles show the maximum distance, $\Delta R_{max}^{ECAL} = 0.05$, considered for the association of hits to a certain track (red stars). Similarly, the calorimeter hits in the fiber hadron calorimeter segment within a $\Delta R_{max}^{HCAL} = 0.30$ are parsed to evaluate possible association to a track.

The total energy of hits associated to a certain track, $i$, is updated at each step exploiting the dual-readout (DRO) corrected and linear response of the calorimeter, i.e.:

\begin{alignat}{3}
\setstretch{2.25}
  & E_{hits\leftrightarrow track,i} \quad && = \quad && \sum_j^N E_{j}^{ECAL} + \sum_k^M E_{k}^{HCAL} \\
  &                                 \quad && = \quad && \frac{\sum S_j - \chi_{ECAL} \sum C_j }{1-\chi_{ECAL}} + \frac{\sum S_k - \chi_{HCAL} \sum C_k}{1-\chi_{HCAL}}
\end{alignat}
in which $S_i$ and $C_i$ are the calibrated calorimeter hits for the scintillation and Cerenkov component, respectively, and $\chi_{ECAL}$, $\chi_{HCAL}$ are the coefficients that describe the difference in the response to electromagnetic and hadronic particles for the homogeneous crystal (ECAL) and sampling fiber (HCAL) segment respectively according to the following Equations:
\begin{align}
\chi_{HCAL} &=\frac{1-(h/e)_s^{HCAL}}{1-(h/e)_c^{HCAL}}\\
\chi_{ECAL} &=\frac{1-(h/e)_s^{ECAL}}{1-(h/e)_c^{ECAL}}
\end{align}
These coefficients were estimated with simulation of single pion events and their values were found to be $\chi_{ECAL} = 0.370$ and $\chi_{HCAL} = 0.445$.

\begin{figure}[!tbp]
    \centering
    \includegraphics[width=0.495\linewidth]{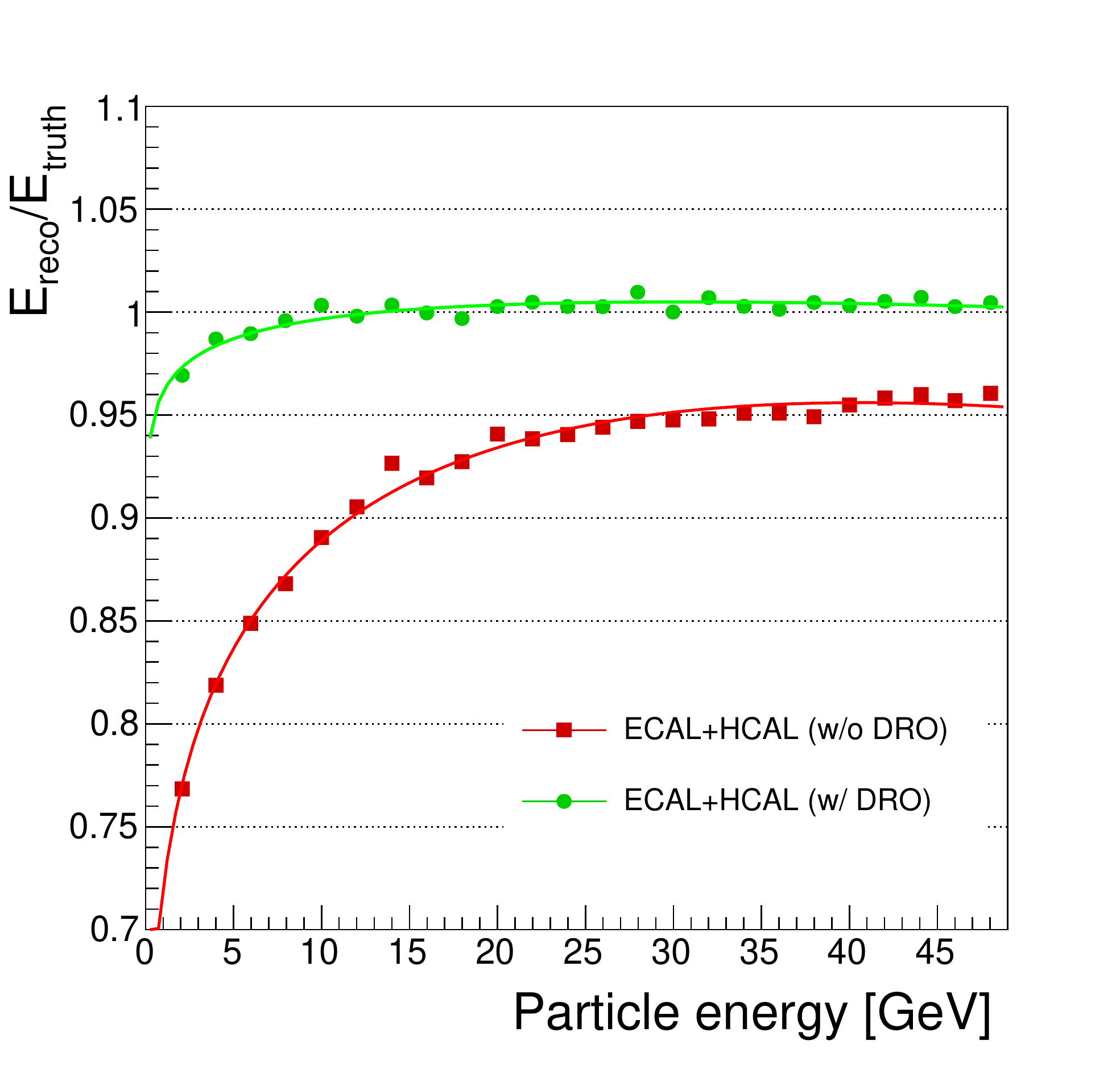}
    \includegraphics[width=0.495\linewidth]{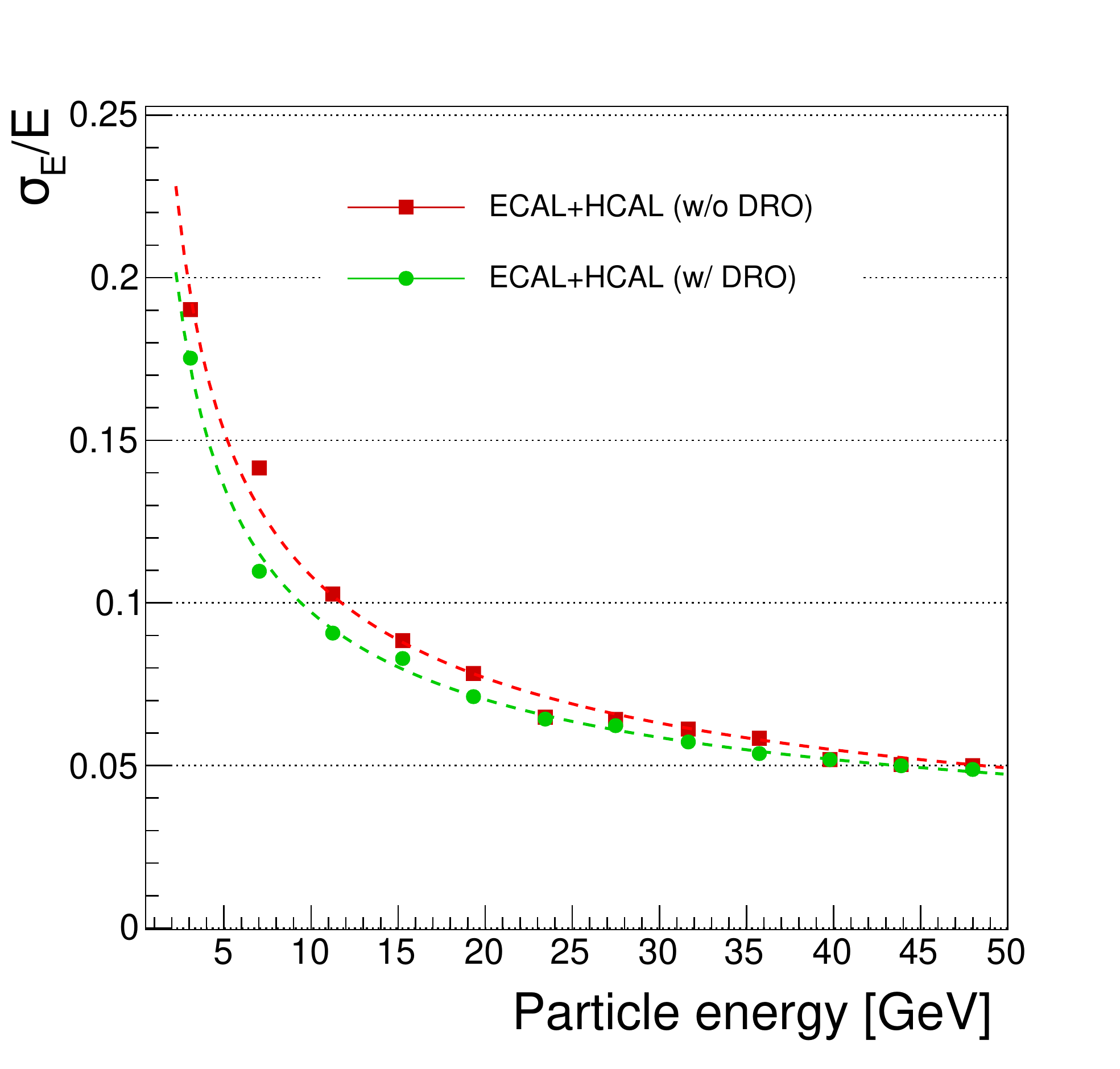}
    \caption{Energy linearity (left) and resolution (right) of the hybrid calorimeter to single pion events in the energy range 1-50 GeV. Red curves correspond to the DRO uncorrected calorimeter response, green points correspond to the DRO corrected response applied on the overall hybrid calorimeter.}
    \label{fig:calorimeter_response_pions}
\end{figure}

\noindent
The calorimeter response (resolution and linearity) to single pions was parameterized using a dedicated simulation of single pion events with variable energy. The results are shown in Fig.~\ref{fig:calorimeter_response_pions}.
As expected the dual-readout approach leads to a rather linear response of the calorimeter to hadrons which combined with an improved energy resolution allows a relatively tight energy matching, thus possibly reducing the impact of the confusion term.
Since some non-linearity (3-5\%) of the calorimeter response for very low energy particles (below 5 GeV) remains after the dual-readout correction, such effect has been taken into account when calculating the expected calorimeter response to a certain track, $E_{target}$.

After each calorimeter hit is added to the candidate cluster, ordered by the angular distance to the track,
the total clustered energy, $E_{hits\leftrightarrow track,i}$, is compared to the expected calorimeter response for that track as shown in the right panel of Fig.~\ref{fig:ppfa_step2}.
If the clustered energy is within an interval of $\pm k_{PFA}\sigma_{E}^{CALO}$ from the expected calorimeter response, i.e. the matching satisfies the Eq.\ref{eq:track_hit_match}, the group of hits is removed from the calorimeter hit collection and the track is added to the PFA collection:
\begin{equation}\label{eq:track_hit_match}
\frac{|E_{hits\leftrightarrow track,i}-E_{target,i}|}{E_{target,i}} < k_{PFA}\cdot\sigma_{E}^{CALO} (E_{target,i})
\end{equation}
where $k_{PFA}$ defines the tightness of the cut around the expected energy and its optimization is discussed later. Conversely, if the clustered hits do not satisfy Eq.~\ref{eq:track_hit_match}, the track is ignored and the calorimeter hits are pushed back into the calorimeter hit collection for the subsequent iterations of track-hit matching.

\begin{figure}[!tbp]
    \centering
    \includegraphics[width=0.495\linewidth]{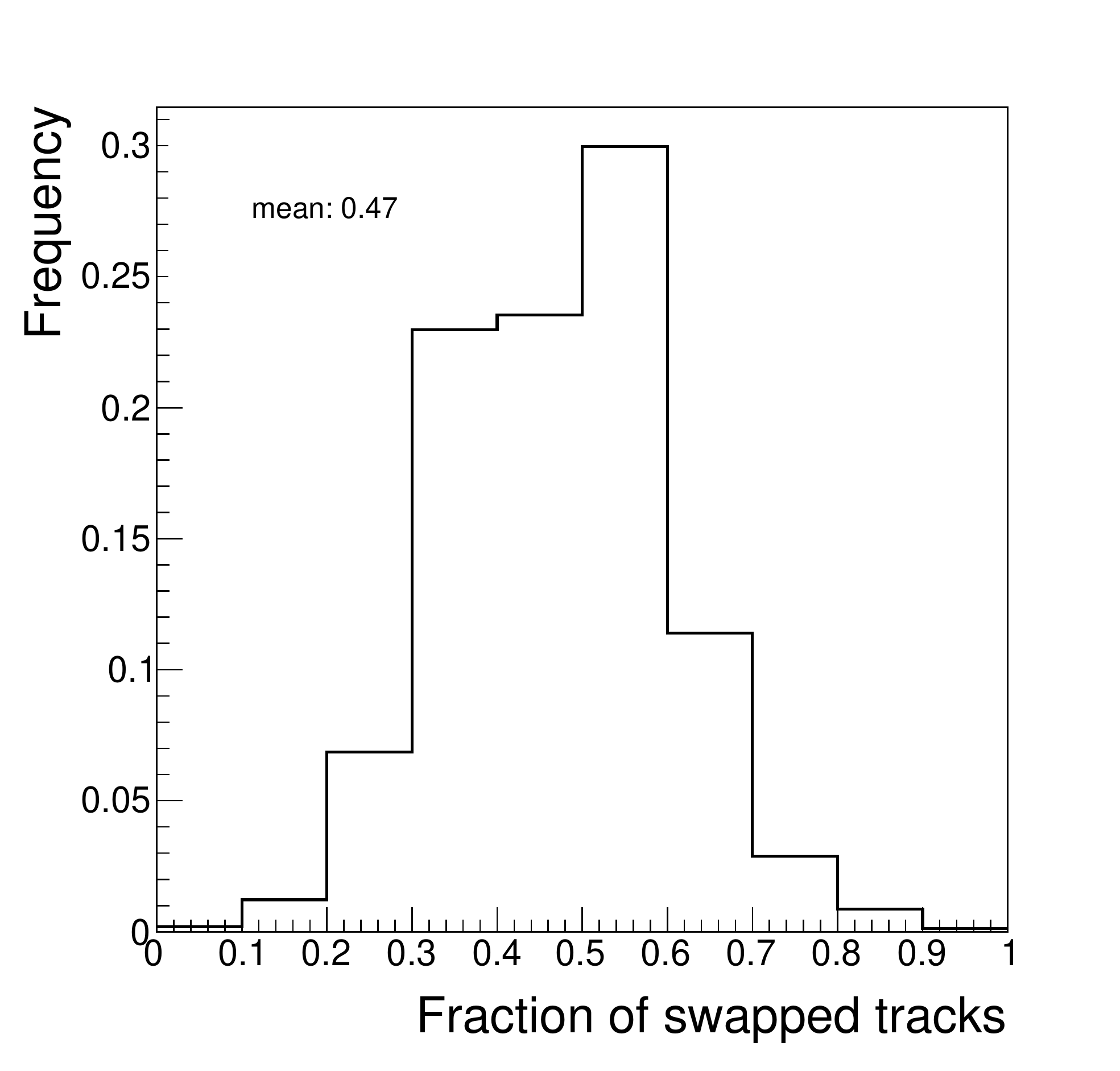}
    \includegraphics[width=0.495\linewidth]{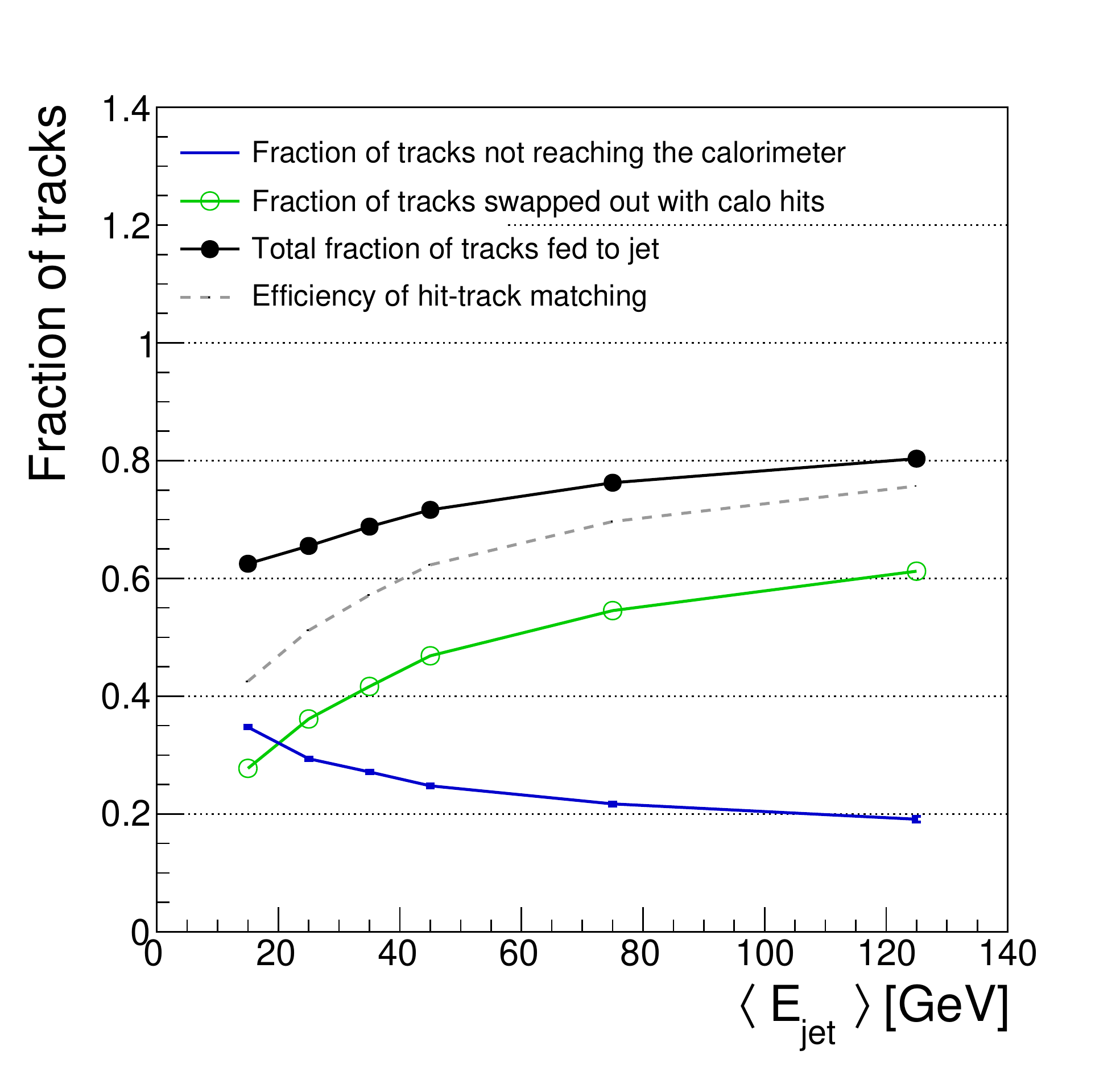}
    \caption{Left: Fraction of charged tracks that are swapped out with calorimeter hits for 45 GeV jets. Right: average fraction of tracks that do not reach the calorimeter (blue), that reach reach the calorimeter and are swapped with calorimeter hits (green dots), fraction of tracks that are fed to the jet clustering algorithm (black dots). The dashed grey curve shows the ratio between the tracks that are successfully matched to calorimeter hits and the tracks that reach the calorimeter and quantifies the efficiency of the algorithm in matching hits to tracks.}
    \label{fig:swapped_track_frac}
\end{figure}

The process continues until all tracks have been parsed, ordered from the highest momentum track to the lowest. The algorithm returns a \emph{PFA collection} containing the charged tracks that have been successfully matched to calorimeter hits (including those tracks that did not reach the calorimeter) and a collection of calorimeter hits that remained unmatched.
As shown in the right panel of Fig.~\ref{fig:swapped_track_frac}, between 60 and 80\% of the tracks (depending on the jet energy) reach the inner surface of the crystal calorimeter and thus take part to the track-hit matching step. Between 40 and 75\% of such tracks have a good enough match to be used to swap out calorimeter hits, and the efficiency for matching grows with the energy of the jet. An example of the distribution of the fraction of tracks which have been successfully matched with calorimeter hits for jets of 45~GeV is reported in the left panel of Fig.~\ref{fig:swapped_track_frac}.
The total fraction of charged tracks that are fed to the jet algorithm (including both, tracks with transverse momentum too small to reach the calorimeter and successfully matched tracks) ranges between 60 and 80\%.

\begin{figure}[!tbp]
    \centering
    \includegraphics[width=0.495\linewidth]{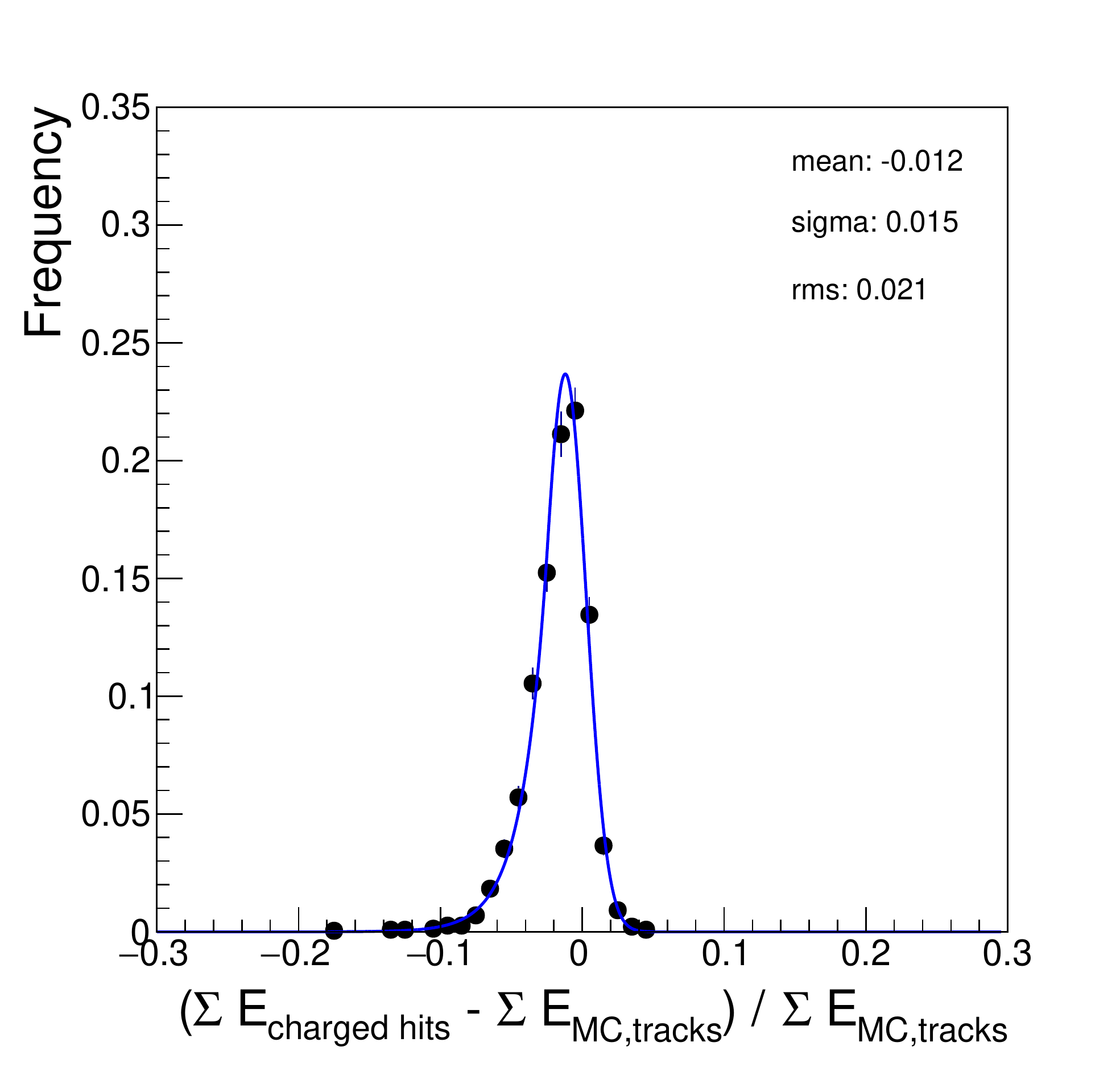}
    \includegraphics[width=0.495\linewidth]{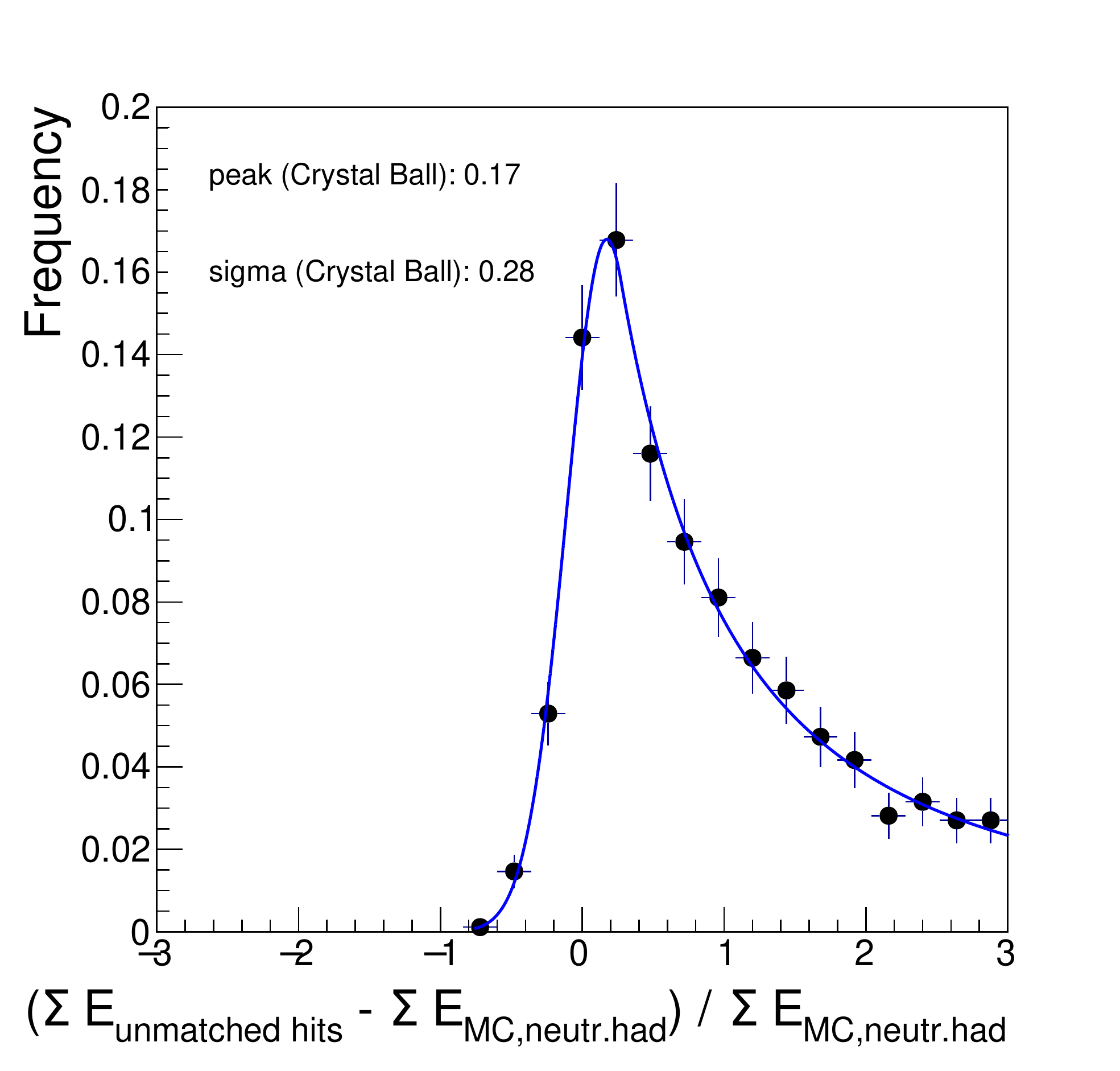}
    \caption{Left: relative residual between the total energy of calorimeter hits associated to charged tracks and the total energy of the corresponding tracks. Right: relative residual between the total energy of calorimeter hits that have not been matched to charged tracks (includes contamination from hits not identified as originating from photons or charged hadrons) and the total energy of neutral hadrons expected from the MC truth.}
    \label{fig:ch_and_neutral_residuals}
\end{figure}

The difference between the total energy of hits associated to charged tracks and expected energy of such tracks according to the MC truth is shown in the left plot of Fig.~\ref{fig:ch_and_neutral_residuals}. As there are many charged hadrons in a jet the overall resolution estimated from a fit with a Crystal Ball function is quite good, although a small tail on the left side is observed. The right plot in Fig.~\ref{fig:ch_and_neutral_residuals} shows instead the difference between the total energy of unmatched hits (hits which were not identified as photons and not assigned to any charged track) and the energy of neutral hadrons in the jet. The distribution peaks at the expectation for the total neutral hadron energy and the tail to the right is due to hits from photons which were not identified by the first step of the algorithm as well as to the fraction of charged hadron hits which have not been swapped out (including those from tracks that were not matched).
When the charged tracks (those that were matched to hits), identified photons and unmatched neutral calorimeter hits are combined into the same jet by the jet clustering algorithm, the overall jet energy distribution is nearly Gaussian with no tails, as reported in the next sections.


The optimal value of $k_{PFA}$ 
was found to be around 1, with small impact on energy resolution if the parameter is varied around this value (0.75-1.25). A larger value of $k_{PFA}$ (2-3) would lead to a larger fraction of tracks matched to a collection of calorimeter hits; however, the fraction of calorimeter hits correctly associated to a certain track would decrease with the risk that the algorithm associates too many hits to the first tracks parsed by the algorithm during the iterative process.


\subsection{Jet clustering algorithm}
A collection of four momentum vectors containing all the unmatched calorimeter hits
and the PFA collection containing the charged tracks is provided as input to the jet clustering algorithm.
In particular the FASTJET package (version 3.3) \cite{fastjetarticle} has been used to cluster the hits and tracks and return the 4-momentum of the primary parton(s). We use the generalized $k_t$ algorithm with $R= 2\pi$ and $p= 1$ designed for $e^{+}e^-$ collision events as provided by the FASTJET package and force the number of jets to two. With these parameters the clustering sequence is identical to the one often referred to as the Durham algorithm \cite{Durham}.
Both scintillation and Cerenkov signals from the crystals and the fiber towers are added to the FASTJET input vector with a label allowing their separation \emph{a posteriori} (once the clustering has been done).

\noindent
For each reconstructed jet the fraction of energy from calorimeter hits is corrected using the typical dual-readout formula and the jet energy is reconstructed as:

\begin{equation}
    E_{jet} = \sum E_{hits,\gamma} + \sum E_{tracks} + \sum E_{hits,unmatched,DRO}~.
\end{equation}

\section{Results}
The algorithm has been tested over a set of samples consisting of off-shell Z boson production decaying to a pair of jets ($e^+e^- \rightarrow Z/\gamma^* \rightarrow jj$) with center-of-mass energies varying from 30 to 250~GeV ($E_{CM}\sim2E_{j,MC}$). Such samples are then used to estimate the energy and angular resolution of jets as a function of jet energy ($E_{j,reco}\sim M_{jj,reco}/2$) in the range 15-125 GeV and compare the calorimeter performance with and without the DR-PFA algorithm. Events containing muons or neutrinos in their final state passed to Geant4 (about 5\% of all the events) are excluded from the study since events with such particles cannot be fully reconstructed using only the calorimeter simulation without a muon detector. To ensure full lateral containment of hadronic showers in the calorimeter both jets are required to be within a pseudorapidity region of $|\eta|<2.0$. In addition, to limit the impact of longitudinal shower leakage due to late starting hadronic showers, it is required that the amount of energy leaked outside the calorimeter is smaller than 1~GeV. A similar cut could be implemented exploiting an external calorimeter layer designed for late shower tagging as the CMS Hadron-Outer calorimeter \cite{CMS_HO}.

In the following we refer to \emph{GenJets} as those obtained by clustering truth-level Monte Carlo (MC) particles that were produced from the hadronization of the partons, simulated by Pythia. \emph{RecoJets} are instead the ones reconstructed from the clustering of the energy deposits of the calorimeter for three main cases: when calorimeter only information and no dual-readout correction is used, when calorimeter only information with dual-readout correction is used, when the dual-readout particle flow algorithm is used (combining calorimeter with tracker information).
The RecoJets are matched to GenJets by minimizing the sum of angular distance between the GenJet-RecoJet pairs in the event.
On an event-by-event basis the relative difference between RecoJets and GenJets is quantified by the following quantities:

\begin{equation}\label{eq:jet_ene}
    \Delta E = \frac{E_{reco} - E_{gen}}{E_{reco}}
\end{equation}
\begin{equation}\label{eq:jet_theta_diff}
    \Delta \theta = \theta_{reco} - \theta_{gen}
\end{equation}
\begin{equation}\label{eq:jet_phi_diff}
    \Delta \phi = \phi_{reco} - \phi_{gen}
\end{equation}

\noindent
The distributions obtained are fitted with a Gaussian function and the jet energy/angular resolutions and the jet energy/angular scales are estimated using the $\sigma$ and $\bar{x}$ resulting from the fit.

\subsection{Jet energy resolution}
To study the contribution of the detector performance to the jet energy resolution we compare the energy reconstructed with different types of \emph{RecoJets} to that of the MC-truth \emph{GenJets}.
The jet energy resolution is estimated by comparing the reconstructed di-jet invariant mass of the system with the one obtained using GenJets (MC truth) according to Equation:

\begin{equation}\label{eq:mjj_residual}
    \frac{M_{jj}^{reco}-M_{jj}^{truth}}{M_{jj}^{truth}}
\end{equation}

\begin{figure}[!tbp]
    \centering
    \includegraphics[width=0.495\linewidth]{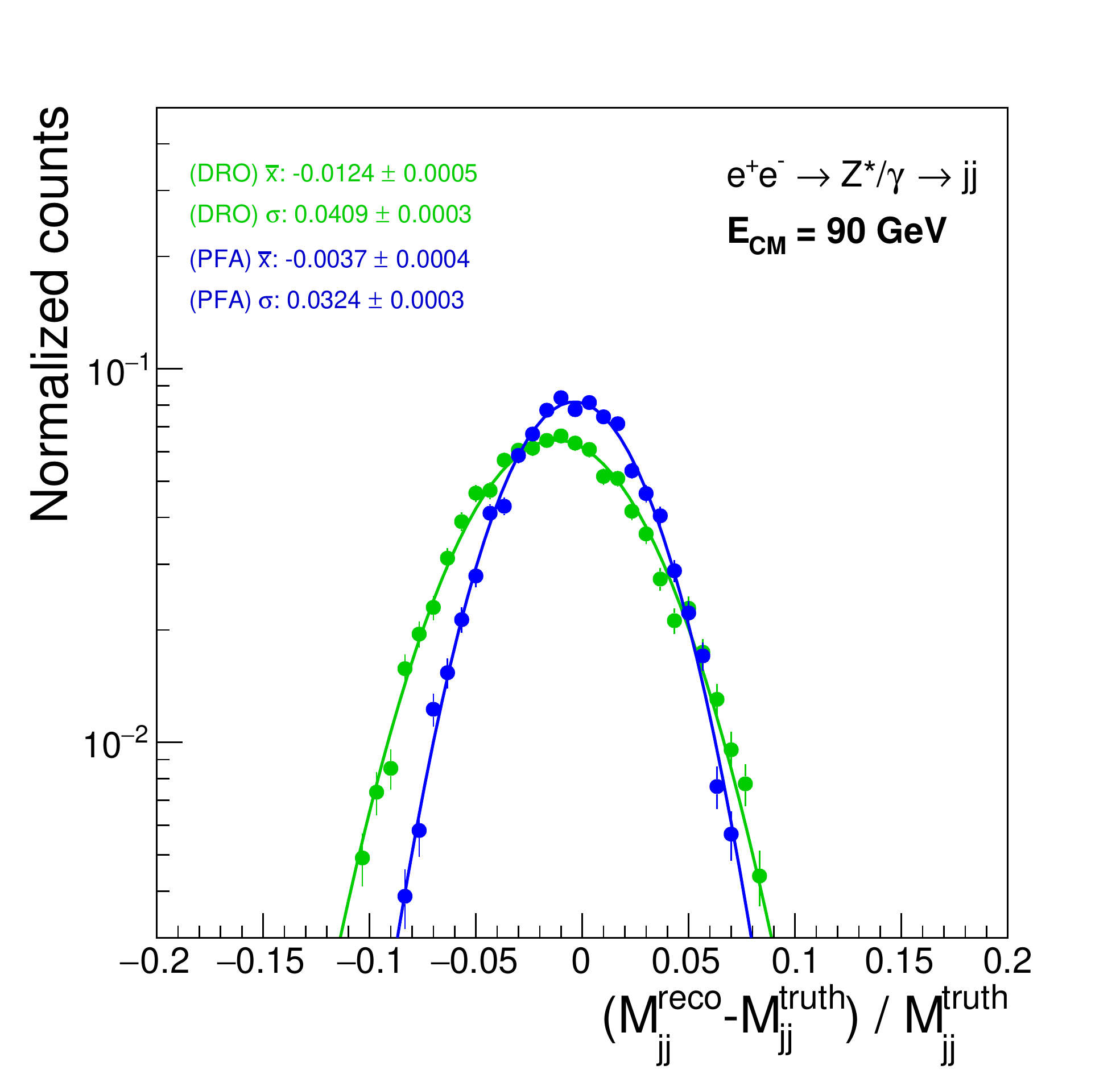}
    \includegraphics[width=0.495\linewidth]{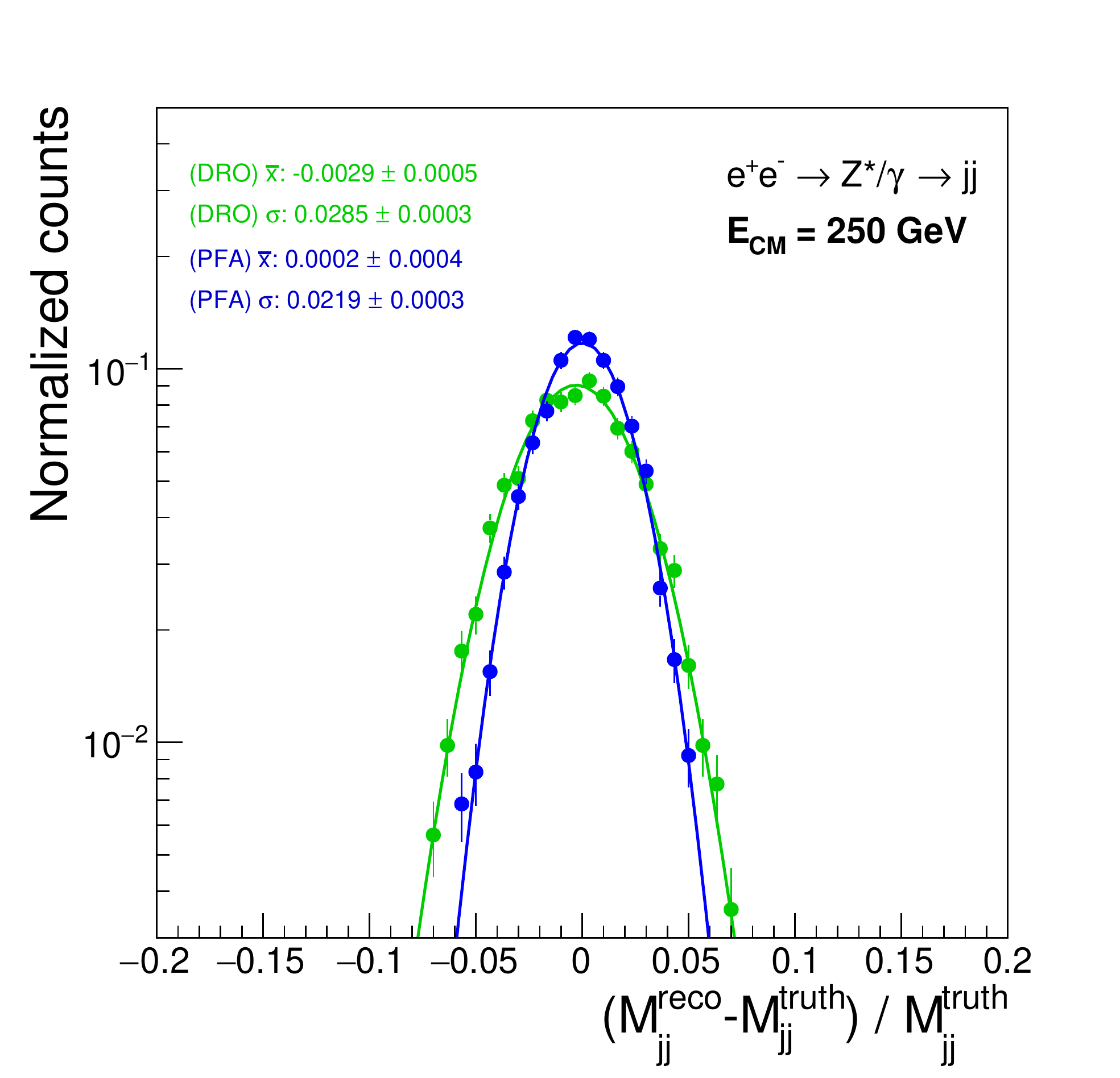}
    \caption{Relative difference in the dijet invariant mass using RecoJets reconstructed with a DRO calorimeter only approach (green) and with the DR-PFA algorithm (blue) with respect to the dijet invariant mass from GenJets. Distributions for a center-of-mass energy of 90 GeV (left) and 250 GeV (right) are shown. Magnetic field of 2~T is used.}
    \label{fig:mjj_res_distribution}
\end{figure}

\noindent
The distributions obtained from Eq.~\ref{eq:mjj_residual} are shown in Fig.~\ref{fig:mjj_res_distribution}. The energy resolution of the invariant mass is estimated as the $\sigma$ obtained from a fit of the distribution using a Gaussian function.
The single jet energy resolution is then estimated by multiplying the $\sigma_{M_{jj}}$ of such distributions by $\sqrt{2}$, as the two jets equally split the energy and contribute to the invariant mass resolution:

\begin{equation}\label{eq:jet_res}
    \sigma_{E_{jet}} = \sigma_{M_{jj}} \cdot \sqrt{2}
\end{equation}

\begin{figure}[!tbp]
    \centering
    \includegraphics[width=0.495\linewidth]{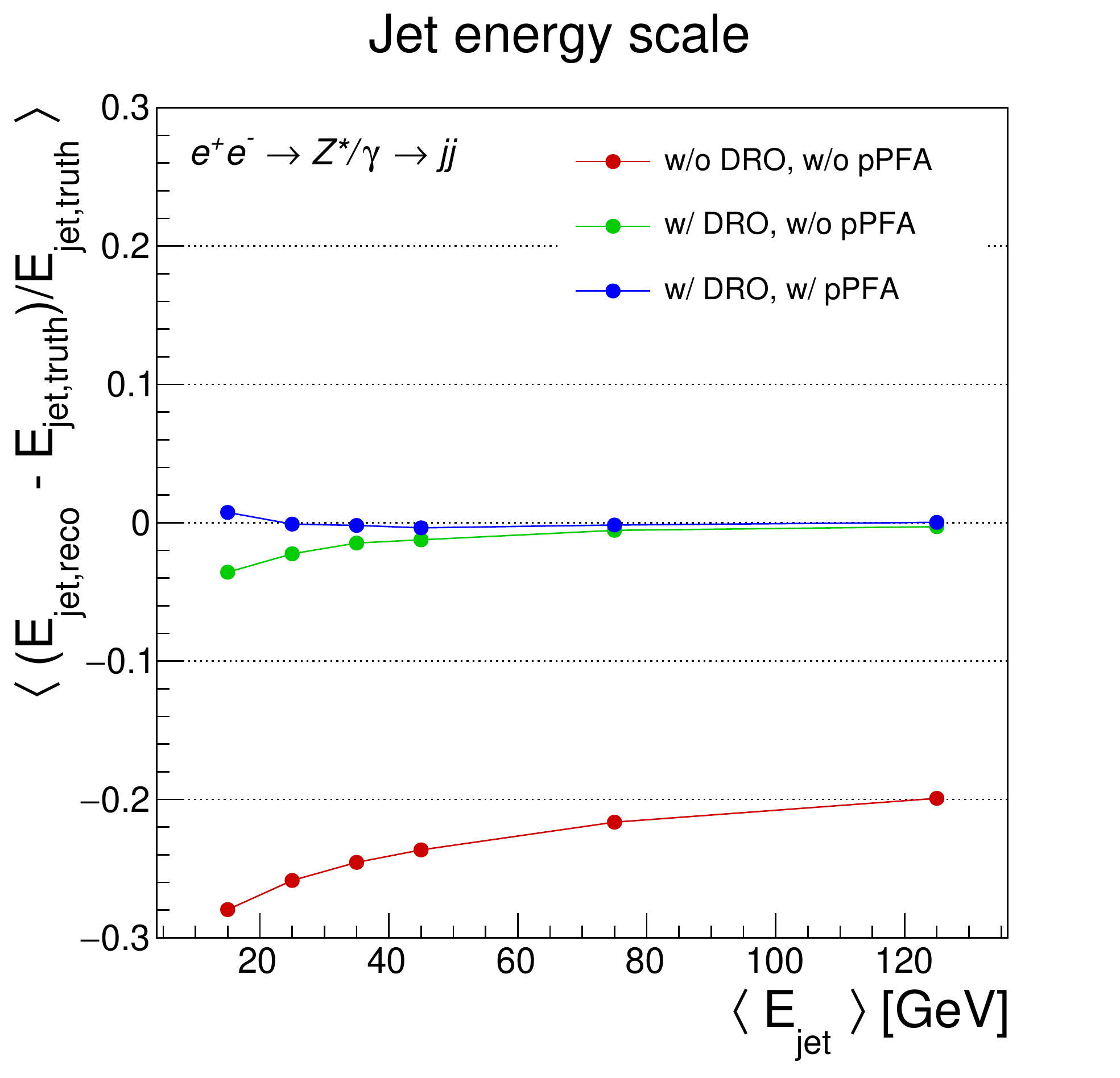}
    \includegraphics[width=0.495\linewidth]{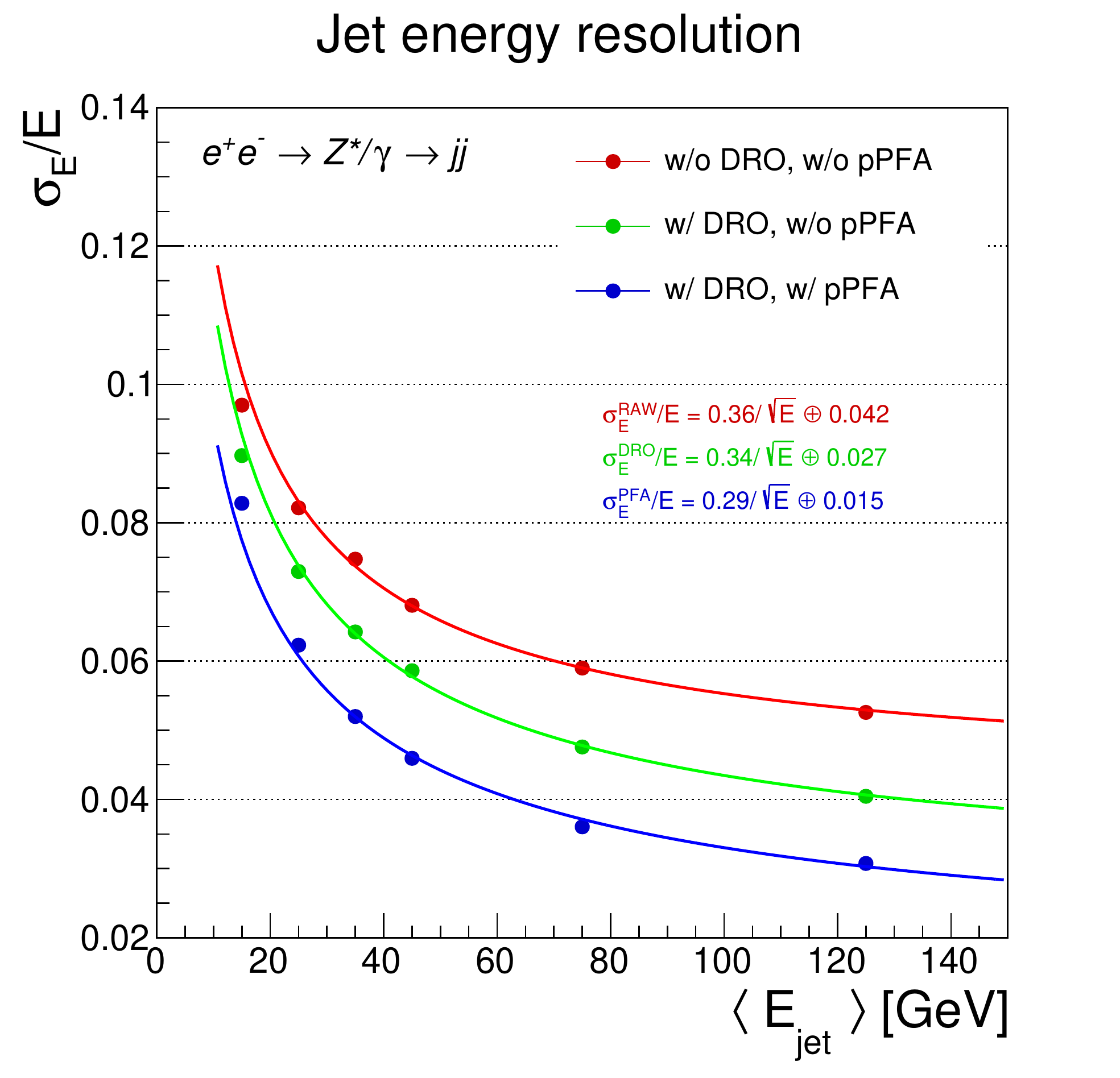}
    \caption{Jet energy scale (left) and resolution (right) of jets as a function of the jet energy for B=2T.}
    \label{fig:jet_lin_res}
\end{figure}

\noindent
The jet energy scale is defined according to Eq.~\ref{eq:jet_ene}.
The results in terms of jet energy scale and resolution are shown as a function of the average jet energy in Fig.~\ref{fig:jet_lin_res}.
As expected, the dual-readout approach restores a linear response of the calorimeter to within $\pm 1\%$ in the energy range from 15 to 125 GeV (with respect to a non dual-readout corrected energy).
An energy resolution of about 4.5\% is achieved for 45 GeV jets when the DR-PFA is used, and decreases down to about 3\% for higher energies. The overall improvement with respect to the `calorimeter-only' performance is substantial. A parameterization of the energy resolution as a function of jet energy (using the sum in quadrature of a stochastic and a constant term) has been performed and the results of the fit are reported in the right plot of Fig.~\ref{fig:jet_lin_res}.

\subsection{Jet angular resolution}
\begin{figure}[!tbp]
    \centering
    \includegraphics[width=0.495\linewidth]{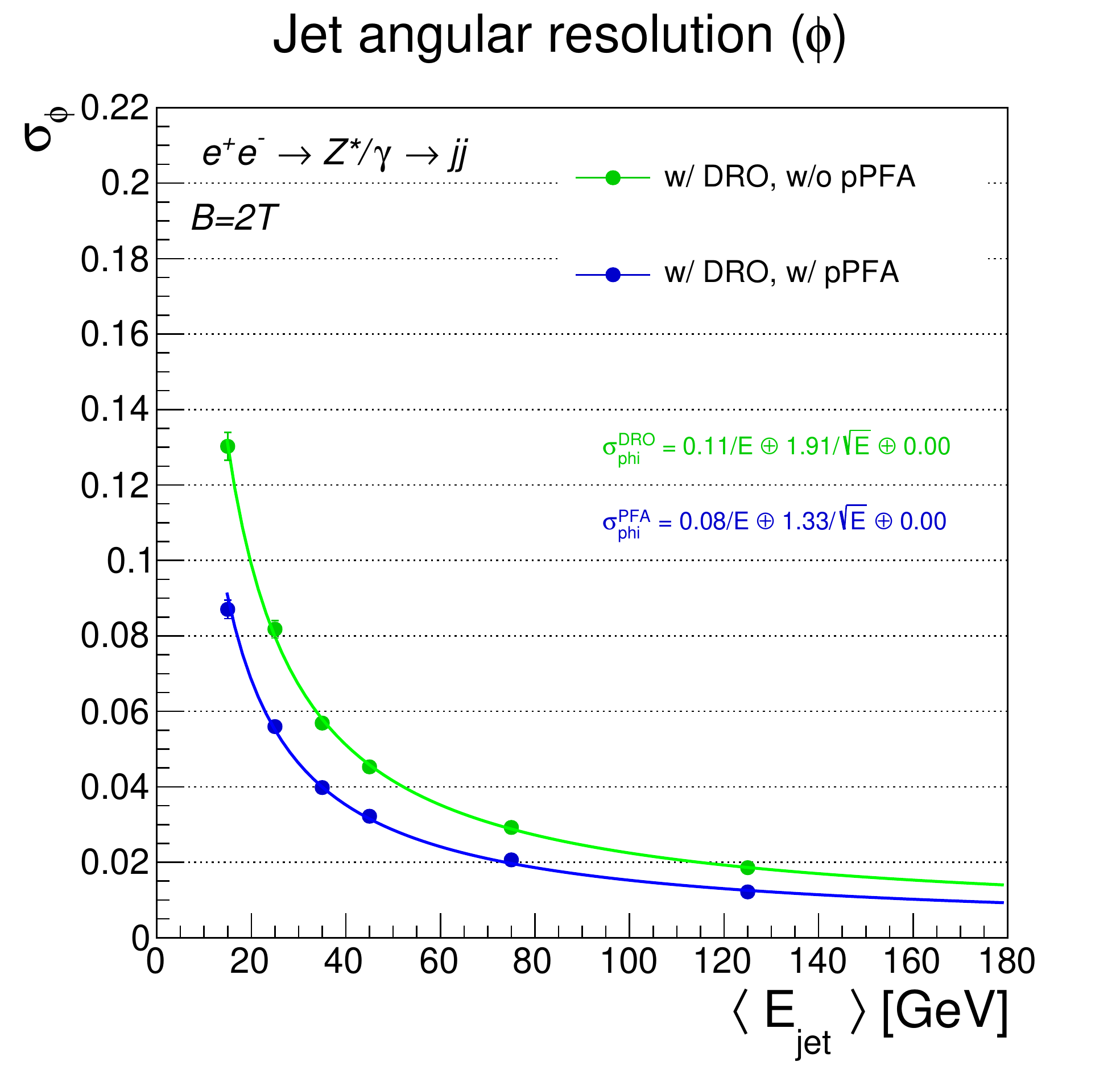}
    \includegraphics[width=0.495\linewidth]{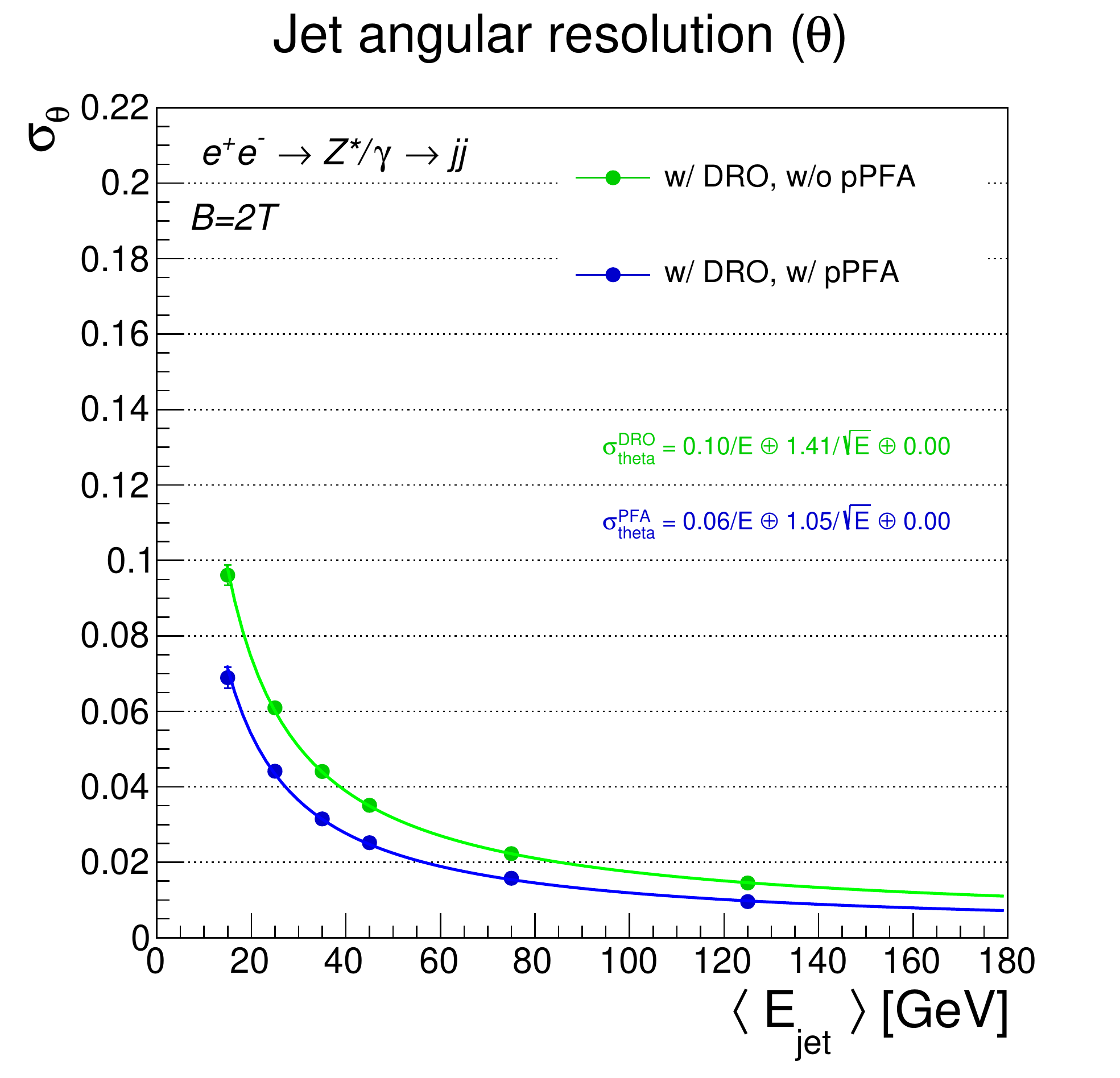}
    \caption{Jet angular resolution along $\phi$ (left) and $\theta$ (right) with 2~T magnetic field as a function of the average jet energy.}
    \label{fig:jet_ang_res}
\end{figure}

The jet angular resolution is estimated from a Gaussian fit of the angular (polar and azimuth angle) difference distribution defined in Eqs.~\ref{eq:jet_theta_diff} and \ref{eq:jet_phi_diff}.
The results are shown in Fig.~\ref{fig:jet_ang_res}. The resolution is improved when the DR-PFA is used and an angular resolution of about 0.01 (0.033) mrad in $\phi$ and 0.008 (0.025) mrad in $\theta$ is obtained for 125 (45) GeV jets. A parameterization of the resolution as a function of jet energy is shown in Fig.~\ref{fig:jet_ang_res}. 

\subsection{Reconstruction of W, Z and H boson invariant masses}
To further validate its performance, the DR-PFA algorithm has been tested over three additional sets of physics events listed below:
\begin{itemize}
    \item $e^+e^- \rightarrow Z H$, $H \rightarrow \chi_1^0\chi_1^0$, $Z \rightarrow jj$\\
    The $Z$ is required to decay only in u,d,s,c quark pairs. The events are cleaned requiring that no neutrinos or muons enter the calorimeter. The same cuts on acceptance and late starting showers discussed in the previous section are applied.
    \item $e^+e^- \rightarrow W^+W^-$, $W^+ \rightarrow \mu \nu$, $W^-\rightarrow jj$\\
    The cleaning requirements are the same as for the $Z$, except that one muon and one neutrino are admitted, and, given the presence of the muon, the requirement on the leakage is modified. The reference leakage is no longer zero, but the energy of the muon, as assumed to be measured in the inner detector, minus the energy it leaves in a cone of $\Delta R = 0.1$ around the muon direction.
    \item $e^+e^- \rightarrow Z H$, $H \rightarrow b \bar b$, $Z \rightarrow \nu \nu$\\
    The cleaning requirements are the same as for the $Z$, except that no neutrinos beyond the ones from the $Z$ decay are accepted. Since the Higgs decay into $b\bar b$ has a large semileptonic branching ratio, the requirement of no neutrinos strongly reduces the statistics. The same cuts on acceptance and late starting shower discussed in the previous section are applied.
\end{itemize}

\begin{figure}[!htbp]
    \centering
    \includegraphics[width=0.495\linewidth]{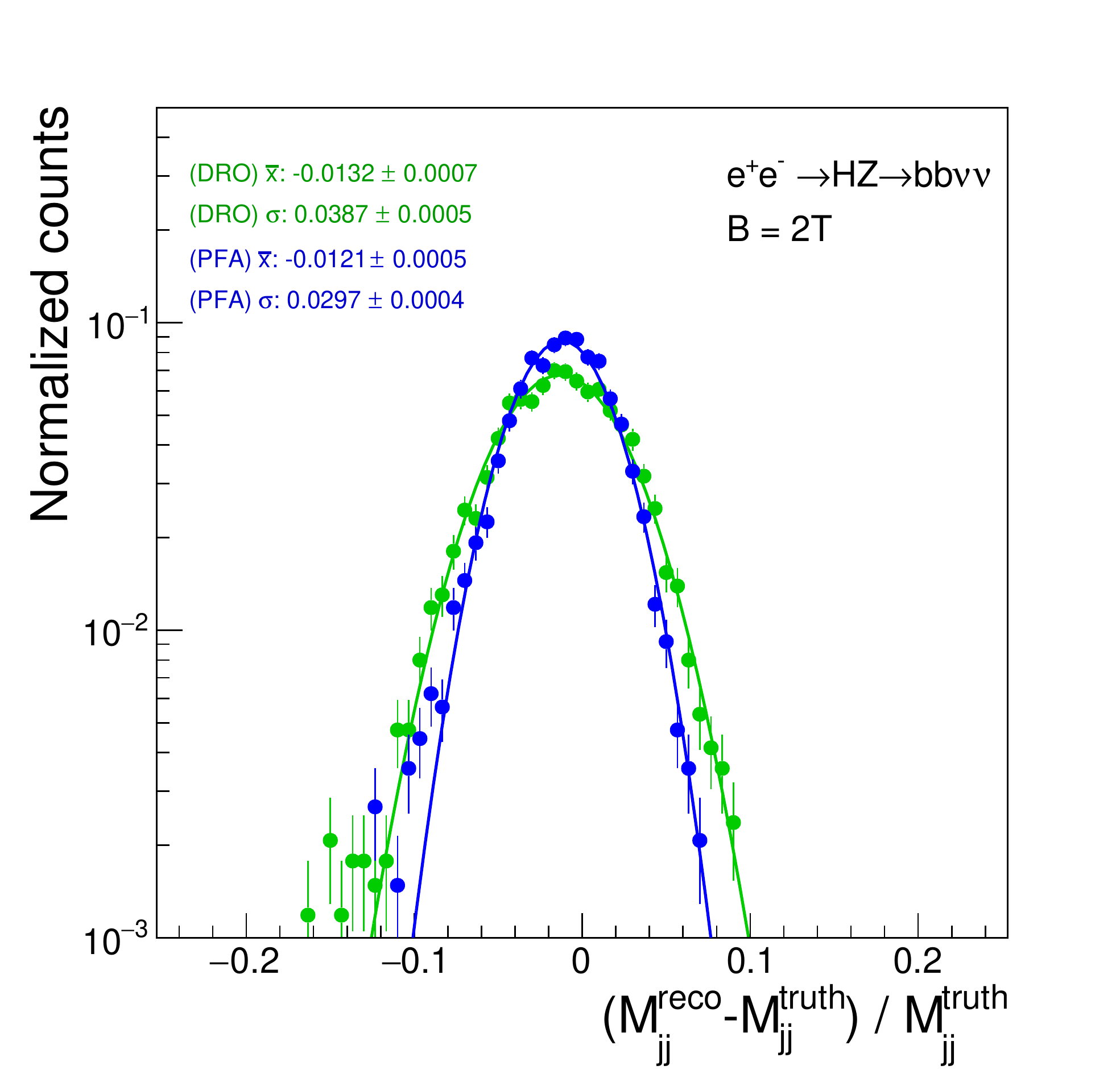}
    \includegraphics[width=0.495\linewidth]{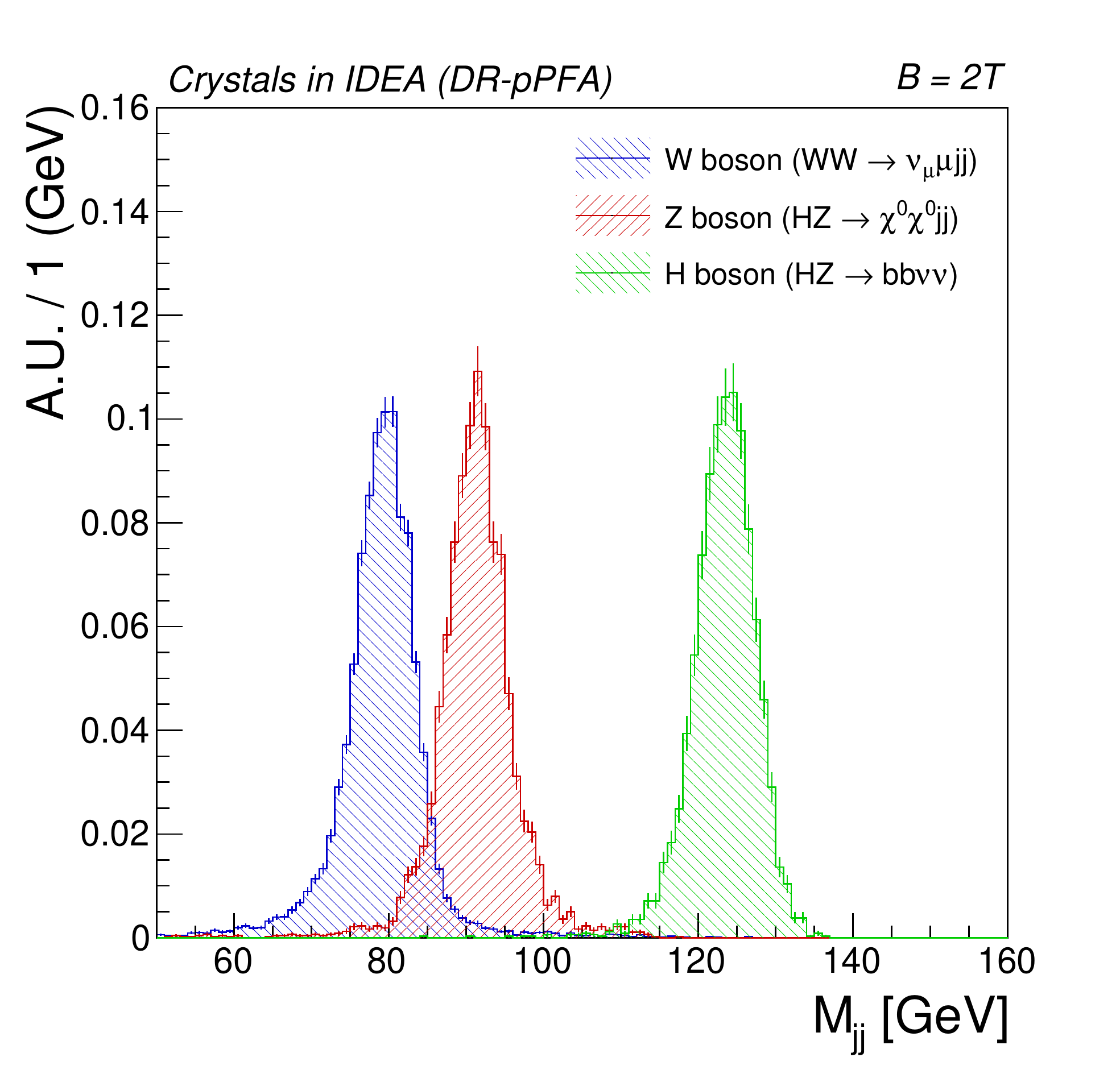}
    \caption{Left: residual of the dijet invariant mass of the H boson reconstructed using the DR-PFA with respect to the MC truth jets. Right: dijet invariant masses of the W, Z and H bosons reconstructed using the DR-PFA algorithm with a magnetic field of 2~T.}
    \label{fig:boson_mass}
\end{figure}

\noindent
An example of the distribution of the relative difference between the H boson dijet invariant mass reconstructed using GenJets and RecoJets is shown in Fig.~\ref{fig:boson_mass}. The use of the DR-PFA algorithm improves the resolution as expected. The dijet invariant mass distributions obtained with the DR-PFA for the W, Z and H bosons are also shown in the right plot of Fig.~\ref{fig:boson_mass}. A good separation between the Z and W boson is achieved. Table~\ref{tab:boson_resolution_compare} reports the energy resolution of the dijet invariant masses for the three bosons with and without the use of the DR-PFA algorithm.



\begin{table}[!htbp]
\centering
\caption{Comparison of dijet invariant mass resolutions for the W, Z and H boson obtained using a 'calorimeter only' approach (with and without dual-readout correction) and using the DR-PFA.}
\vspace{0.2cm}
\begin{tabular}{l|c|c|c}
\hline
              & 'Calo only' (w/o DRO) & 'Calo only' (w/ DRO) & DR-PFA (w/ DRO)      \\ \hline\hline
W boson       & 5.6\%                  & 4.7\%                  & 3.8\%     \\
Z boson       & 4.7\%                  & 4.1\%                  & 3.3\%     \\ 
H boson       & 4.5\%                  & 3.9\%                  & 3.0\%     \\ \hline
\end{tabular} 
\label{tab:boson_resolution_compare}
\end{table}


\section{Discussion and outlook}\label{sec:discussion}
The algorithm developed and tested in this study shows the potential of integrating the dual-readout information and excellent energy measurements of photons and neutral hadrons from a hybrid segmented calorimeter with the particle flow approach of exploiting measurements of charged particle momenta from a tracking system.
Future studies will aim at reproducing these results with a more complete detector geometry that includes a real tracking system.

The algorithm is at an early stage and further improvements can be sought in every step of the algorithm described in Section~\ref{sec:dr_pfa}.
The excellent electromagnetic energy resolution ($3\%/\sqrt{E}$) of the crystal calorimeter segment offers the possibility to cluster $\pi^0$ photons ahead of the jet clustering algorithm and could reduce scrambling of photons across neighboring jets in 4 and 6 jets event topologies as discussed in \cite{Lucchini_2020}.
The identification of neutral seeds used in this algorithm is very simple while more sophisticated techniques that exploit  neural networks for clustering have been recently developed for instance by the CMS ECAL detector \cite{CMS-DP-2021-032} and could be exploited.

While these two aspects could further improve the initial clean up of neutral calorimeter hits, even larger improvement could come from a revision of the track-hit matching step where a global optimization of the matching, e.g. exploiting graph-theory techniques, capable of considering all possible tracks and hit combinations as whole rather than association via an iterative procedure.
In addition, the excellent time resolution, on the order of 20 ps, that the crystal calorimeter segments could provide for minimum ionizing particles and electromagnetic showers would offer an additional handle for event clean up and possibly enhance the selection of calorimeter hits that should be clustered into jets from other delayed backgrounds, especially if charged tracks carried independent timing information from tracking layers.

At the same time, the consolidation of the present algorithm, would also open the possibility to explore a DR-PFA oriented optimization of the detector design.
In particular, the transverse and longitudinal segmentation of the detector could be fine tuned and optimal radial envelopes for the tracking system, solenoid, and calorimeter segments could be found. 
For instance, the full exploitation of the fiber granularity with respect to the HCAL granularity considered in this study could enhance the performance of future particle flow algorithms, with or without a machine learning based approach.

\section{Summary}\label{sec:summary}
A prototype particle flow algorithm (DR-PFA) which exploits the dual-readout information and excellent energy resolution in a hybrid segmented calorimeter was developed and tested with simulated dijet events from e$^+$e$^-$ collisions. An energy resolution of about 4.5\% is achieved for 45 GeV jets allowing for a good separation of the Z and W boson invariant masses.
The DR-PFA algorithm also improves the jet angular resolution to below 0.01 mrad for 125 GeV jets.
While the algorithm is at an early stage of development it shows the potential of exploiting dual-readout information and precise photon measurements to achieve excellent jet energy resolution in a calorimeter with moderate longitudinal segmentation. Future work in this direction should target an increase in the sophistication of the algorithm (e.g. exploiting for instance neural networks, graph-theory approaches and the exploitation of the full granularity of the DR fiber HCAL) to improve the matching of calorimeter hits and charged tracks for an enhanced integration of the particle flow and the dual-readout approaches.

\acknowledgments
The authors would like to acknowledge the IDEA calorimeter group for supporting this work with comments and suggestions.


\bibliography{mybibfile}




\end{document}